\documentclass[aps,prd,twocolumn,showpacs,superscriptaddress,a4paper,floatfix]{revtex4}
\usepackage{graphicx}
\usepackage{longtable}
\usepackage{appendix}
\usepackage[centertags]{amsmath}
\usepackage{color}
\usepackage[latin1]{inputenc}
\usepackage[english]{babel}
\usepackage{footnote}
\usepackage[stable]{footmisc}
\usepackage{slashbox}
\usepackage{xspace}
\usepackage{verbatim}

\begin{document}

\title[Higher order Laguerre--Gauss mode degeneracy in realistic, high finesse cavities]
{Higher order Laguerre--Gauss mode degeneracy in realistic, high finesse cavities}

\author{Charlotte Bond}
\affiliation{School of Physics and Astronomy, University of
Birmingham, Edgbaston, Birmingham B15 2TT, UK}
\author{Paul Fulda}
\affiliation{School of Physics and Astronomy, University of
Birmingham, Edgbaston, Birmingham B15 2TT, UK}
\author{Ludovico Carbone}
\affiliation{School of Physics and Astronomy, University of
Birmingham, Edgbaston, Birmingham B15 2TT, UK}
\author{Keiko Kokeyama}
\affiliation{School of Physics and Astronomy, University of
Birmingham, Edgbaston, Birmingham B15 2TT, UK}
\author{Andreas Freise}
\affiliation{School of Physics and Astronomy, University of
Birmingham, Edgbaston, Birmingham B15 2TT, UK}

\date{\today}

\begin{abstract} 	
Higher order Laguerre--Gauss (LG) beams have been proposed for use in future 
gravitational wave detectors, such as upgrades to the Advanced LIGO detectors  and 
the Einstein Telescope, for their potential to reduce the effects of the 
thermal noise of the test masses.  This paper
details the theoretical analysis and simulation work carried out to investigate the 
behaviour of LG beams in realistic optical setups, in particular the coupling between
different LG modes in a linear cavity. We present a new analytical approximation
to compute the coupling between modes, using Zernike polynomials to describe
mirror surface distortions.
 We apply this method in a study of the behaviour of the LG$_{33}$ mode within realistic arm cavities, 
using measured mirror surface maps from the Advanced LIGO project. 
We show mode distortions that can be expected to arise due to the degeneracy of 
higher order spatial modes within such cavities and relate this to the theoretical analysis.  
Finally we identify the mirror distortions which cause significant coupling from the LG$_{33}$ mode 
into other order 9 modes and derive requirements for the mirror surfaces.
\end{abstract}

\maketitle

\section{introduction}
The sensitivities of second generation gravitational wave detectors such as Advanced LIGO and Advanced Virgo
are expected to be limited by the thermal noise of the test masses within a significant range of signal frequencies around 100 Hz~\cite{Rowan05}.  
To reach even better sensitivities, it has 
been proposed to use laser beams with an intensity pattern other than that of the fundamental Gaussian
beam to reduce the effects of this thermal noise~\cite{Vinet07, Mours06}.  A beam whose 
intensity is distributed more homogeneously over the mirror surface, for the same clipping losses, benefits
from a more effective averaging over the mirror surface distortions caused by thermal effects~\cite{Vinet09}.  
The specific advantage of using 
higher order LG modes, as opposed to mesa~\cite{Ambrosio03} and conical~\cite{Bondarescu08} beams, is that they 
are compatible with spherical mirrors as currently used in GW detectors and other high precision optical setups.  
Research into the potential of the LG$_{33}$ mode in gravitational wave detectors has been carried out using 
numerical simulations and table-top experiments~\cite{Chelkowski09, Fulda10}.
The sensing and control signals for an  LG$_{33}$ beam were found to perform as well as for the fundamental mode 
in all aspects examined and the LG$_{33}$ behaved as expected in short linear and triangular optical cavities.
However, an optical cavity resonant for a higher--order Gaussian mode is degenerate so that a number of modes
can resonate at the same time. This is a fundamental difference to a well designed cavity for the fundamental 
Gaussian mode, in which any resonant enhancement of other modes can be suppressed. This degeneracy
can potentially cause additional optical losses. Simulations have shown that 
the use of the LG$_{33}$ beam, compared to the fundamental mode, LG$_{00}$, could result in a significant contrast 
defect at the dark fringe~\cite{Miller,Galimberti,Yamamoto}.
It is the aim of this paper to investigate how mirror surface distortions affect the purity of an LG$_{33}$ beam in high 
finesse cavities, by analytical calculation and numerical simulation.  We will focus on the direct coupling from a distorted mirror 
and how this affects the mode content in a linear cavity, with our final aim to produce specifications for the mirror surfaces.

\section{Representing mirror surface distortions and LG modes}
\subsection{Mirror surface maps}
\label{sec:mirror_maps}
 
\begin{figure}[b]
\begin{center}
\includegraphics[scale=0.26]{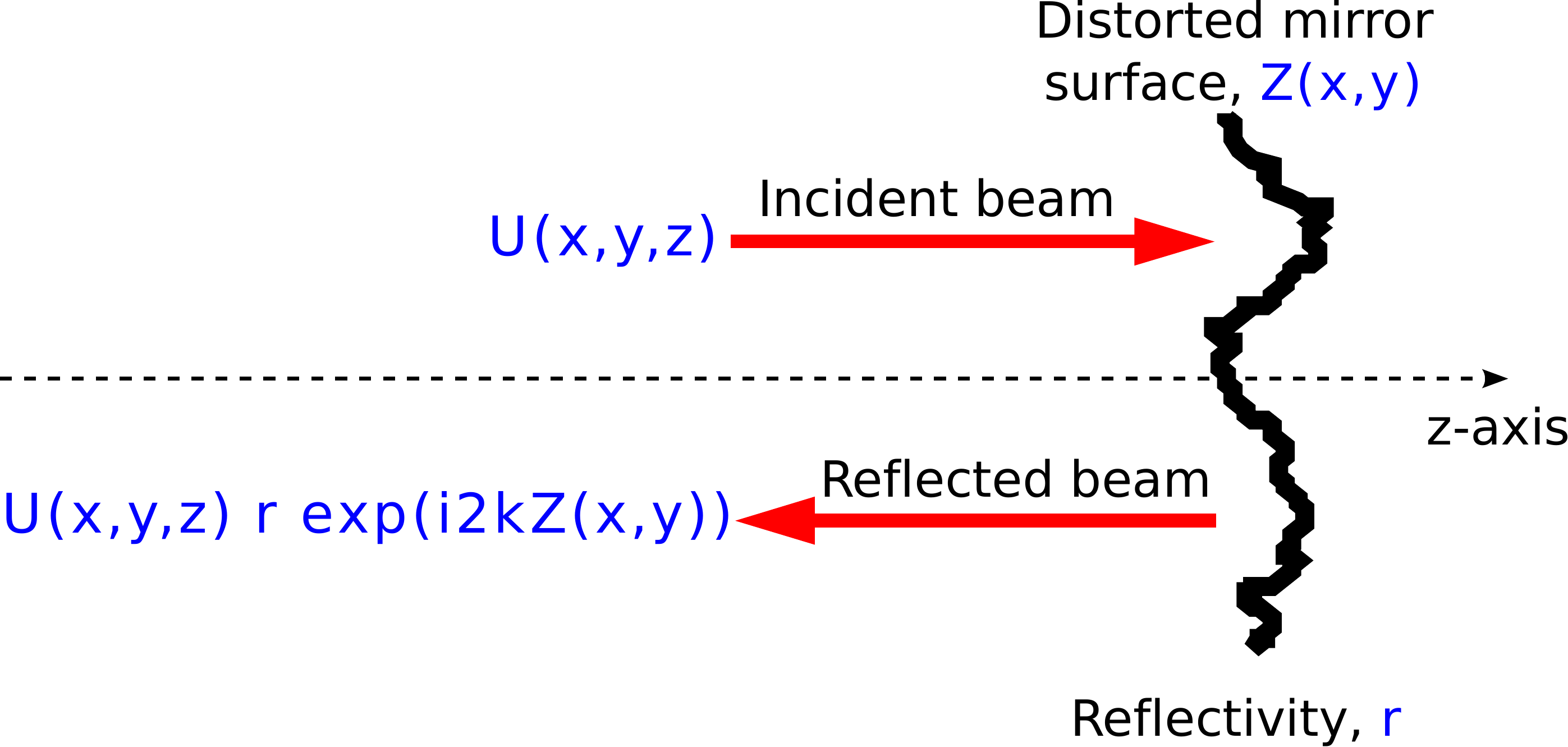}
\end{center}
\caption{A diagram illustrating the phase shift introduced when a beam is reflected from a distorted mirror surface with reflectivity $r$.
The surface $Z(x,y)$ is defined as a height field across a plane perpendicular to the optical axis along $z$.}
\label{fig:coupling_diagram}
\end{figure}

Ideally the mirrors in gravitational wave detectors should be perfectly smooth with a radius of curvature 
matching that of the incident beam.  However, real mirrors deviate from a perfect surface, altering the 
beams which interact with them.  
If a beam $U(x,y,z)$ is incident on a distorted surface described by $Z(x,y)$ and uniform 
reflectivity $r$, then the reflected beam is given by:
\begin{equation}
U_{ref}(x,y,z)=U(x,y,z) \ r  \exp{(i2kZ(x,y))},
\end{equation}
Fig.~\ref{fig:coupling_diagram} illustrates this effect.  

\begin{figure}[t]
   \begin{center}
   \includegraphics[scale=0.53]{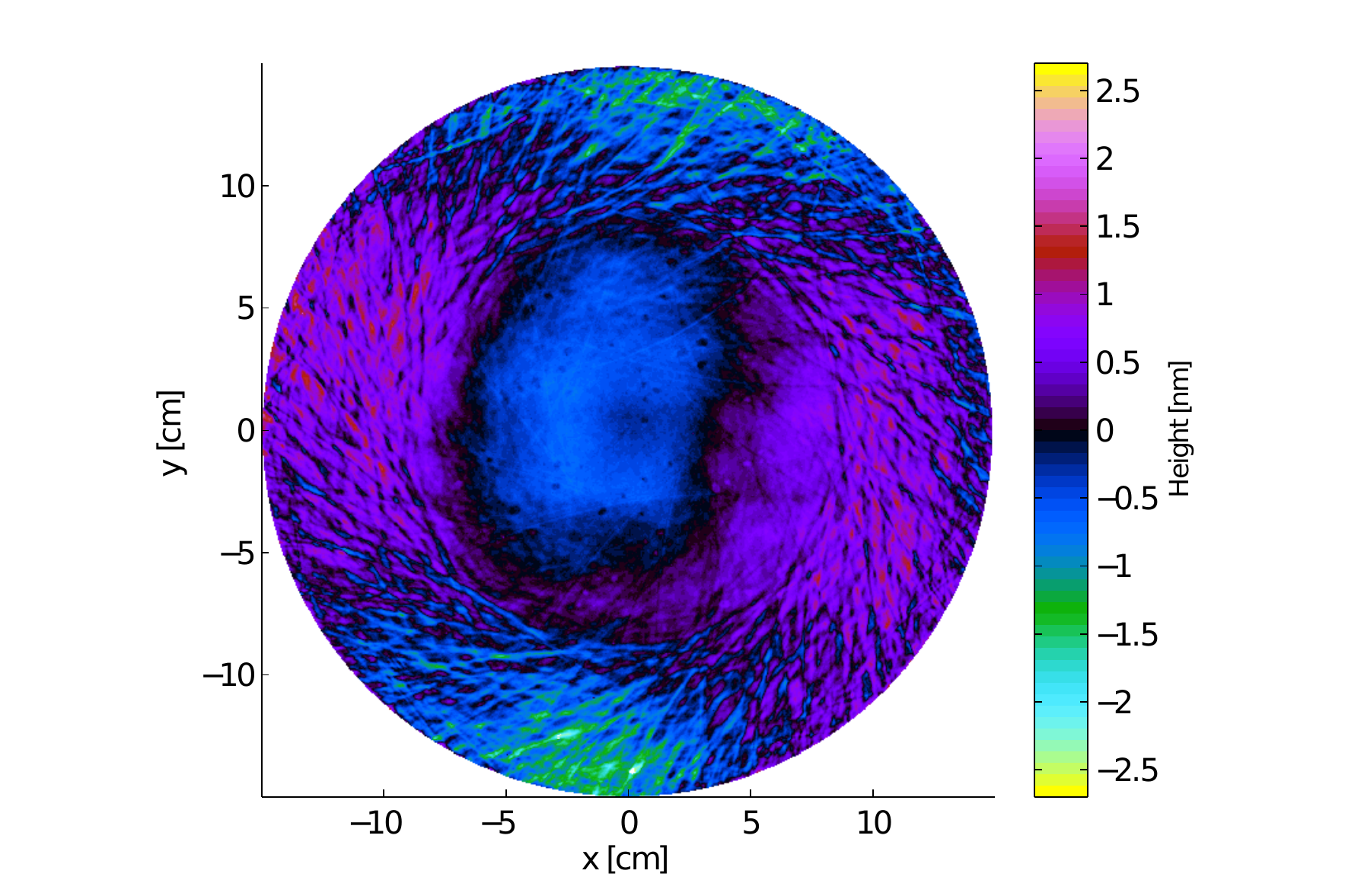}
   \end{center}
   \caption{A surface plot of the mirror map ETM08 corresponding to the surface heights of 
   an Advanced LIGO end test mass.  The curvature has been fitted and removed from the 
   map data.}
   \label{fig:ETM08}
\end{figure}

In order to investigate the effects of surface distortions measured mirror surface maps can be used.  The term \emph{mirror map} refers 
to an array of data detailing the optical properties of a mirror, often its surface height in nanometers. 
This data can be used to represent realistic mirrors in numerical simulations of gravitational wave detectors. 
Mirror maps have been produced from uncoated Advanced LIGO mirror substrates which represent the best mirror surfaces of this kind currently available.  
In the following we have made use of one such map, the surface map of the substrate ETM08~\cite{ETM08}.
The deviation of this map surface from a perfectly spherical surface with radius of curvature
2249.28\,m is shown in Fig.~\ref{fig:ETM08}.
This substrate shows an RMS surface figure error of 0.523\,nm.

\subsection{Zernike polynomials}
\begin{figure}[bhtp]
\begin{minipage}[c]{0.1\textwidth} 
\end{minipage}
\begin{minipage}[c]{0.5\textwidth}
\centering
\includegraphics[scale=0.1]{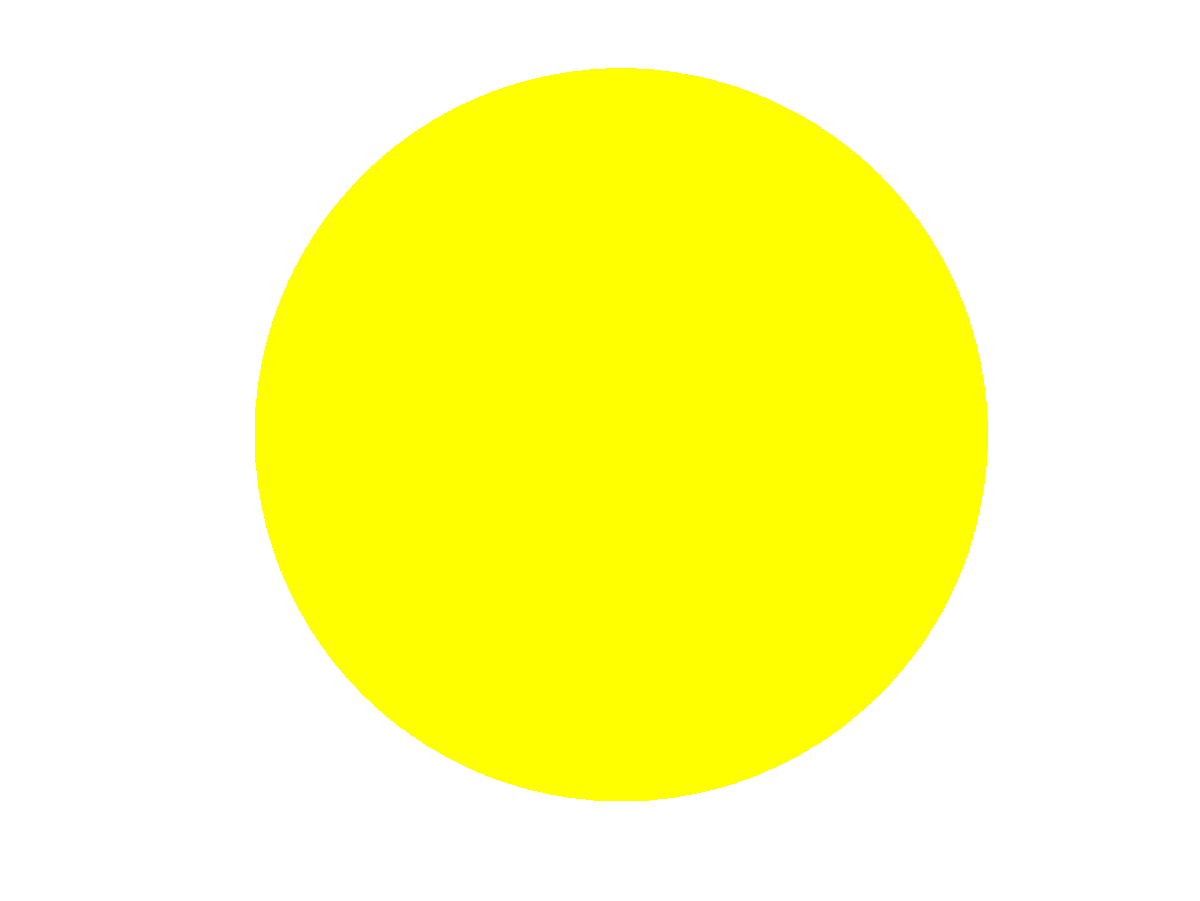}
\\
$n=0$
\end{minipage}
\begin{minipage}[c]{0.1\textwidth}
\end{minipage}
\begin{minipage}[c]{0.5\textwidth}
\centering
\includegraphics[scale=0.1]{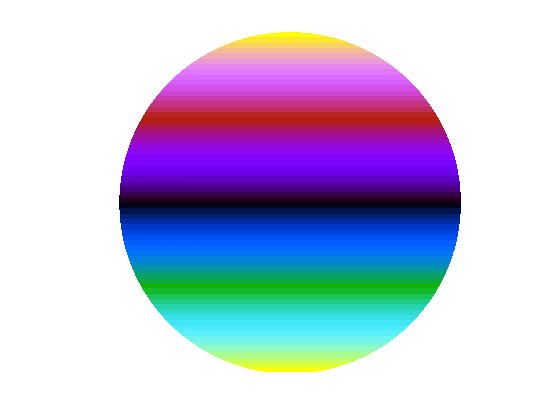}
\includegraphics[scale=0.1]{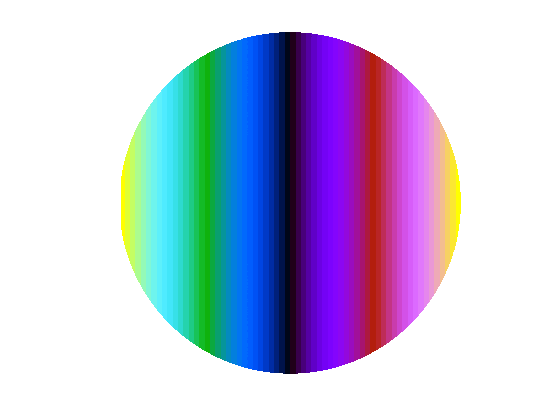}
\\
$n=1$
\end{minipage}
\begin{minipage}[c]{0.1\textwidth}
\end{minipage}
\begin{minipage}[c]{0.5\textwidth}
\centering
\includegraphics[scale=0.1]{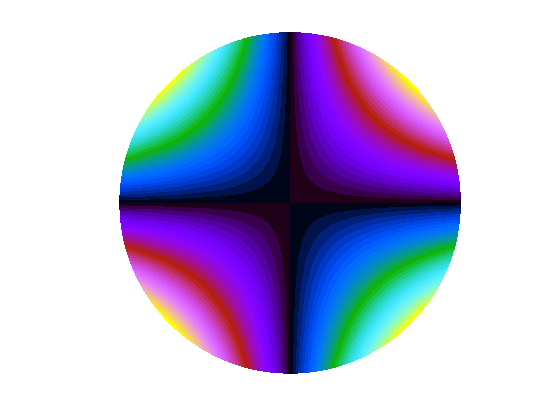}
\includegraphics[scale=0.1]{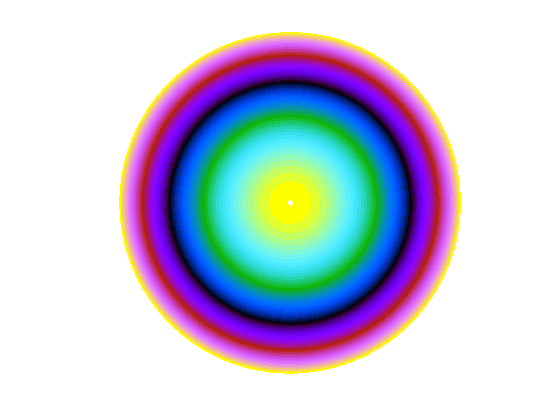}
\includegraphics[scale=0.1]{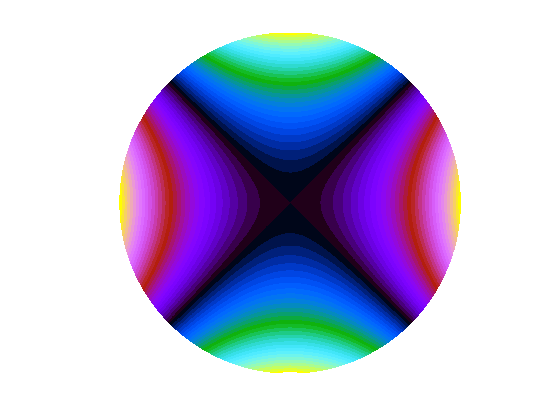}
\\
$n=2$
\end{minipage}
\begin{minipage}[c]{0.1\textwidth}
\end{minipage}
\begin{minipage}[c]{0.5\textwidth}
\centering
\includegraphics[scale=0.1]{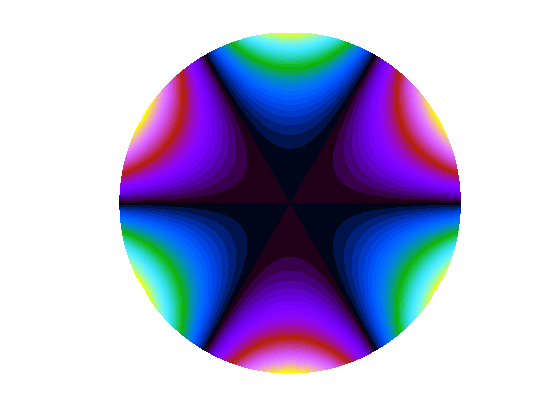}
\includegraphics[scale=0.1]{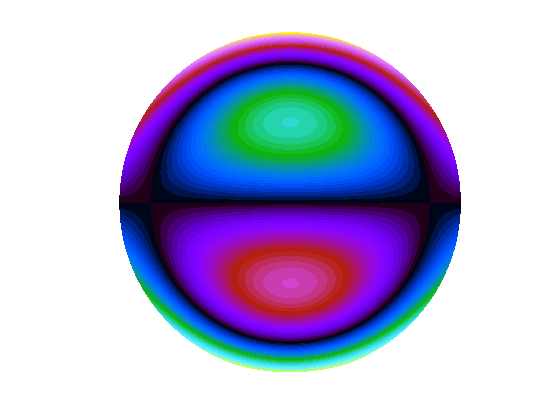}
\includegraphics[scale=0.1]{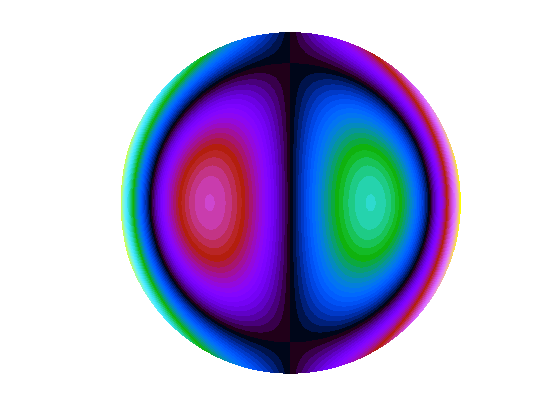}
\includegraphics[scale=0.1]{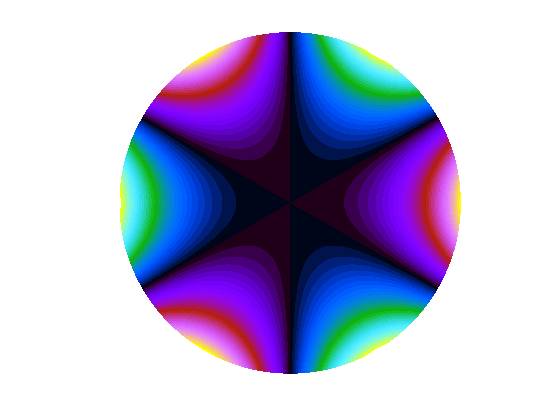}
\\
$n=3$
\end{minipage}
\begin{minipage}[c]{0.1\textwidth}
\end{minipage}
\begin{minipage}[c]{0.5\textwidth}
\centering
\includegraphics[scale=0.1]{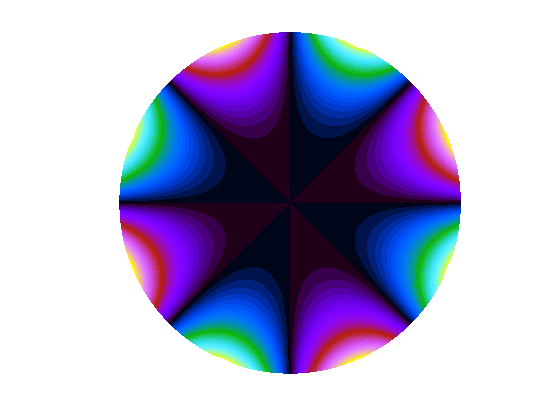}
\includegraphics[scale=0.1]{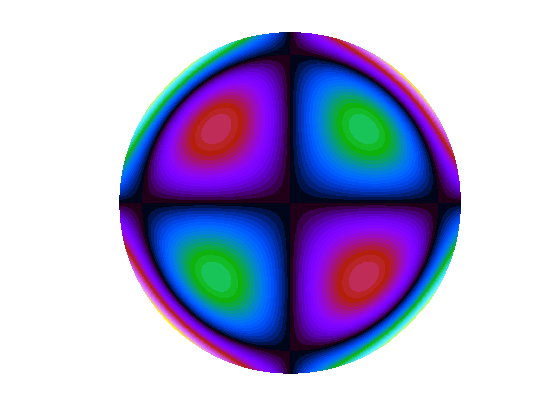}
\includegraphics[scale=0.1]{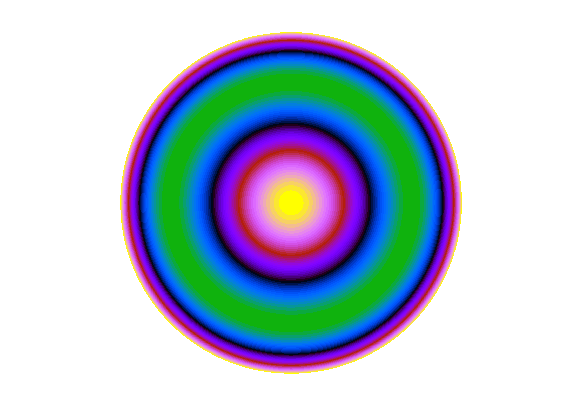}
\includegraphics[scale=0.1]{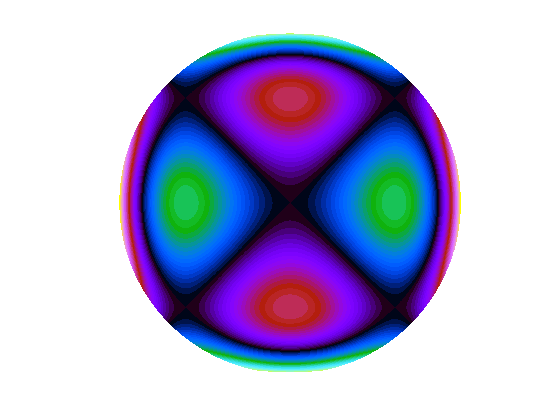}
\includegraphics[scale=0.1]{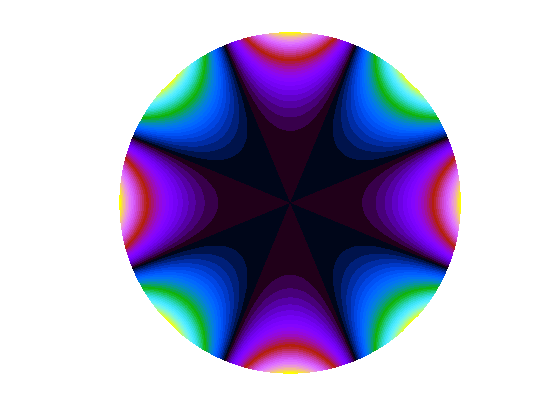}
\\
$n=4$
\end{minipage}
\caption{Plots of the non-zero Zernike polynomials from $n=0$ to $n=4$ with the odd polynomials with $m=-n$ on the far left and the even polynomials with $m=n$ on the far right, in steps of 2.  The colour scale represents negative surface heights with greens and blues, zero with black and positive surface heights with reds and purples.}
\label{fig:zern}
\end{figure}

Zernike polynomials are well suited for the purposes of describing mirror surface distortions.  
Zernike polynomials can be used to describe classical 
distortions such as tilts and curvatures~\cite{Zernike}.  They are a complete set of functions which are orthogonal over 
the unit disc and defined by radial index, $n$, and azimuthal index, $m$, with $ n \geq m \geq 0$.  For any index $m$ we have
one odd and one even polynomial~\cite{Wolf}:
\small
\begin{equation}
   \begin{array}{ll}
       Z_{n}^{+m}(\rho,\phi)=  A_{n}^{+m} \ \cos(m\phi)R_{n}^{m}(\rho) & \mbox{even polynomial} \\
       \\
         Z_{n}^{-m}(\rho,\phi)= A_{n}^{-m} \ \sin(m\phi)R_{n}^{m}(\rho) & \mbox{odd polynomial}\\
   \end{array}
\end{equation}
\normalsize
where $\rho$ is the normalised radial coordinate, $\phi$ is the azimuthal angle, $A_{n}^{\pm m}$ is the amplitude and $R_{n}^{m}(\rho)$ is the radial function.  The radial function is given by the following sum:
\small 
\begin{equation}
R_{n}^{m}(\rho)=  
   \sum_{h=0}^{\frac{1}{2}(n-m)} \frac{(-1)^{h}(n-h)!}{h! \left(\frac{1}{2}(n+m)-h\right)! \left(\frac{1}{2}(n-m)-h\right)!} \rho^{n-2h} 
\end{equation}
\normalsize
for $n-m$ even and 0 otherwise.  This gives $n+1$ non-zero Zernike polynomials for each value of $n$ (for $m=0$ the odd polynomial is zero).  Fig.~\ref{fig:zern} shows the surfaces described by the Zernike polynomials corresponding to orders (n) 0 to 4.  The lower order polynomials 
represent some common optical distortions, some of which are summarised in table~\ref{table:zernike}.  

\begin{table}[b]
\begin{center}
\begin{tabular}{ccc}
\hline
\hline
$n$ & $m$ & Common name \\
\hline
0 & 0 & Offset \\
1 & $\pm1$ & Tilt in $x$/$y$ direction \\
2 & 0 & Curvature \\
2 & $\pm2$ & Astigmatism \\
3 & $\pm 1$ & Coma along $x$/$y$ axis \\
\hline
\hline
\end{tabular}
\end{center}
\caption{Summary of some common names for the lower order Zernike polynomials~\cite{Zernike}.}
\label{table:zernike}
\end{table}

The odd polynomial describes the same surface as the even polynomial, but rotated by 90$^{\circ}$.  
Combinations of the odd and even polynomials relate to this same distortion rotated by a given angle.  
The magnitude of this surface distortion can be given by the root mean squared amplitude of the polynomials:
\begin{equation}
A_{n}^{m}=\sqrt{\left(A_{n}^{-m} \right)^{2}+ \left(A_{n}^{+m}\right)^{2}}
\end{equation}
where $+$ refers to the even polynomial and $-$ refers to the odd polynomial.  Any surface defined over a disc can be described by a sum of Zernike polynomials, with the higher order polynomials representing the higher spatial frequencies present in the surface. 

\subsection{Laguerre-Gauss modes}

The shape of any paraxial beam can be described as a sum of Hermite-Gauss or Laguerre-Gauss modes. 
The Laguerre-Gauss modes are a complete and orthogonal set of functions
defined by radial index $p$ and azimuthal index $l$. The helical type of LG modes are typically given as~\cite{Lasers}:
\begin{equation}      
\begin{split} 
   U_{p,l}(r,\phi,z)={}& \frac{1}{w(z)}\sqrt{\frac{2p!}{\pi(|l|+p)!}}\exp{\left(i\left(2p+|l|+1\right)\Psi(z)\right)} \\
      		{}& \times \left(\frac{\sqrt{2}r}{w(z)}\right)^{|l|} L^{|l|}_{p}\left(\frac{2r^{2}}{w^{2}(z)}\right) \\
		{}& \exp{\left(-\frac{ikr^{2}}{2R_{c}(z)}-\frac{r^{2}}{w^{2}(z)}+il\phi\right)}
\end{split}
\end{equation}
where $k$ is the wavenumber, $w(z)$ is the beam spot 
size parameter, $\Psi(z)$ is the 
Gouy phase and $R_{c}(z)$ is the radius of curvature of the 
beam.  $L^{|l|}_{p}(x)$ refer to the associated Laguerre 
polynomials.  When considering these beams in cavities we note 
that the resonance conditions for these 
beams differ from that of a plane wave due to the $(2p+|l|+1)\Psi(z)$ 
phase shift.  The order of an LG mode is 
given by $ 2p+|l|$ and modes with the same order will acquire the 
same round trip 
phase shift whilst circulating in a cavity.  Therefore the cavity is 
degenerate for LG modes of the same order.

The effects of mirror surface distortions on the shape of a reflected 
beam can be described in terms of coupling between LG modes.  
When a perfectly aligned Gaussian beam is reflected by a perfectly 
spherical mirror with the radius of curvature of the mirror
matching that of the beam's phase front, the shape of the reflected beam is 
identical to the shape of the incident beam, or, in
other words, the mode composition has not changed.
However, if the mirror surface is distorted the reflected beam will generally 
have a different mode composition.  
The coupling from an incident mode (indices $p$ and $l$) impinging on a completely reflecting surface $Z$ into a 
mode  (indices $p'$ and $l'$) in the reflected beam can be described by a coupling coefficient~\cite{Bayer-Helms, Freise10}:

\begin{equation}
\label{eq:k_int}
k_{p,l,p',l'}^{Z}=\int_{S}U_{p,l}\exp{\left(2ikZ\left(r,\phi\right)\right)}U^{*}_{p',l'}
\end{equation}
$Z$ describes the surface height of the mirror and $S$ describes an infinite plane perpendicular to the optical axis.  

\begin{figure*}[htbp]
\begin{center}
\includegraphics[scale=0.45]{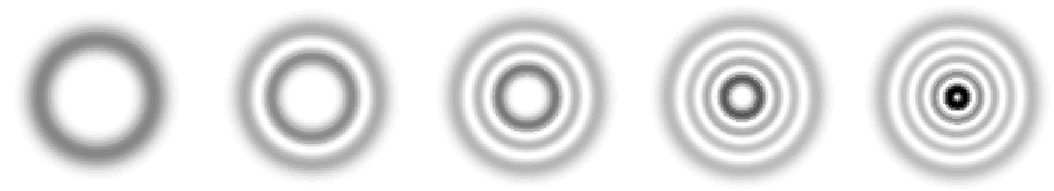}
\end{center}
\caption{Plots of the intensity distributions of the order 9 helical Laguerre--Gauss modes.  From left to right: LG$_{0\pm9}$, LG$_{1\pm7}$, LG$_{2\pm5}$, LG$_{3\pm3}$ and LG$_{4\pm1}$.}
\label{fig:LG9}
\end{figure*}

Currently the fundamental mode, LG$_{00}$, is used in gravitational wave detectors.  Investigations have shown
 that mirror surface distortions have little effect on the beam purity when LG$_{00}$ is used.  
 The presence
 of mirror surface distortions introduces modes into the detectors other 
 than the input mode, but since LG$_{00}$ is the only
 mode of order 0 any coupling out this mode will result in modes 
 of a different order, which will be suppressed in the cavities.  
The LG$_{33}$ mode is one of several modes of order 9.  In total there are 10 order 9 modes; LG$_{0,\pm9}$, LG$_{1,\pm7}$, 
LG$_{2,\pm5}$, LG$_{3,\pm3}$ and LG$_{4,\pm1}$ (Fig.~\ref{fig:LG9}).  
These modes will be resonant in the arm cavities of the detectors, potentially resulting in a large proportion of the circulating 
power being in modes other than LG$_{33}$.  
The distortions 
present in the mirrors in each arm cavity will be different and so the mode content in each arm will differ, resulting in a larger
contrast defect at the main beam splitter. 

\section{Analytical description of mode coupling via mirror surface distortions}
\label{sec:analysis}
Using Zernike polynomials as a description of mirror surface distortions we can look at the coupling between Laguerre--Gauss modes analytically. 
The coupling between different Laguerre-Gauss modes when the surface is described by a particular Zernike polynomial is given by:
\begin{equation}
k^{n,m}_{p,l,p',l'}=\int_{S}U_{p,l}\exp{\left(2ikZ_{n}^{m}\right)}U^{*}_{p',l'}
\label{eq:original_eq}
\end{equation}
In order to simplify the integral we use the fact that when $kZ$ is small we can approximate:
\begin{equation*}
\exp{(2 i  k Z)}\approx 1+2 i k Z
\end{equation*}
The amplitudes of the Zernike polynomials in the mirrors used in gravitational wave detectors are not expected to exceed 10 nm.  With a wavelength of 1064 nm we have $2 k Z\approx0.1$ and so the approximation should be suitable for this investigation.  We are concerned with coupling into other modes, not back into the input mode, so the equation to solve becomes:
\begin{equation}
k^{n,m}_{p,l,p',l'}=\int_{0}^{2\pi} \int_{0}^{R} U_{p,l}U^{*}_{p',l'}(2ikZ_{n}^{m})r dr d\phi
\end{equation} 
due to the orthogonal properties of LG modes.  

Both the Zernike polynomials and the Laguerre-Gauss modes can be easily separated into their angular and radial parts.  The 
angular function to integrate is:
\begin{equation*}
\exp{(i\phi(l-l'))}     \begin{array}{ll} 
      \cos{(m\phi)} & \ \ \ \mbox{even polynomial}\\
      \sin{(m\phi)} & \ \ \ \mbox{odd polynomial}\\
   \end{array}
\end{equation*}
Considering the even Zernike polynomial we obtain:
 \begin{equation*}
    \begin{array}{c}
       \displaystyle\int_{0}^{2\pi}e^{i\phi(l-l')} \frac{e^{im\phi}+e^{-im\phi}}{2} d\phi  =  \\
       \\
            \displaystyle\left[ \frac{e^{i\phi(l-l'+m)}}{2i(l-l'+m)} + \frac{e^{i\phi(l-l'-m)}}{2i(l-l'-m)} \right]^{2\pi}_{0} \\
    \end{array}
 \end{equation*}
As the integral is evaluated over the entire unit disc and $e^{i0}=e^{iN\times2\pi}=1$, where $N$ is an integer, the integral is equal to 0.  The only combination of Zernike polynomials and Laguerre-Gauss beams to give a non-zero result occurs when one of the exponentials disappears before the integration takes place.  This occurs when we have:
\begin{equation}
m=|l-l'|
\label{eq:m_condition}
\end{equation}
The same condition also gives the only non-zero results for the odd Zernike polynomials.  This is a very interesting result as it suggests that surfaces described by Zernike polynomials will only cause significant coupling from one LG mode to another if the Zernike azimuthal index is equal to the difference between the azimuthal indices of the two modes.  This requirement for $m$ also gives the minimum order ($n$) of Zernike polynomial required to produce significant coupling, as $m \leq n$.   

Using this condition we can integrate with respect to $\phi$.  The integrals were found to be $\pi$ for the even Zernike polynomials and $\pm i\pi$ for the odd polynomials, depending on the sign of $(l-l')$.   
The final equation is given by:
\begin{equation}
   \begin{split}
            k_{p,l,p',l'}^{n,m} = {}& A_{n}^{m} k \sqrt{p!p'!(p+|l|)!(p'+|l'|)!} \\
         {}&  \times \left| \sum_{i=0}^{p} \sum_{j=0}^{p'} \sum_{h=0}^{\frac{1}{2}(n-m)}  \frac{(-1)^{i+j+h}} {(p-i)!(|l|+i)! i!}  \right. \\
         {}& \times \frac{1}{(p'-j)!(|l'|+j)!j!} \frac{(n-h)!}{(\frac{1}{2}(n+m)-h)!} \\
         {}& \left. \times  \frac{1}{ \left(\frac{1}{2}(n-m)-h\right)! h!} \frac{1}{X^{\frac{1}{2}(n-2h)}} \right. \\           {}& \left. \times \ \gamma (i+j-h+\frac{1}{2}(|l|+|l'|+n)+1,X) \ \right|  \\
   \end{split}
   \label{eq:k_eq}
\end{equation}
where $X=\frac{2R^{2}}{w^{2}}$, $R$ is the Zernike radius and $w$ is the beam radius, and $\gamma$ is the lower incomplete gamma function.  The full derivation is given in Appendix~\ref{sec:aderivation}.

In our approximation of the coupling coefficients we only consider the magnitude of the coefficients.  However, the real coefficients have both real and imaginary parts indicating that there is some phase shift caused by the distortions.  Therefore, when considering the coupling from a surface in terms of the coupling from the individual polynomials making up the surface we also need to consider the phase shifts.  The largest possible coupling from a surface occurs when all the individual Zernike couplings have the same phase and therefore the magnitude of the coupling is equal to the sum of the individual couplings.

\section{Analysis of coupling into order 9 modes}

We  want to verify the results of the analytical description of the coupling coefficients.  Using the condition for significant coupling outlined previously in Sec.~\ref{sec:analysis} we can identify the azimuthal Zernike indices which will cause a large amount of coupling from LG$_{33}$ into the other order 9 modes.  These are summarised in table~\ref{table:m_for_order_9}.  Because $m \leq n$ this condition for the azimuthal index also tells us the lowest order ($n$) Zernike polynomial required to cause a large amount of coupling from the LG$_{33}$ beam into each of the other order 9 modes.
\begin{table}[b]
\begin{center}
\begin{tabular}{lccccccccc}
\hline
\hline
$m$ & 2 & 2 & 4 & 4 & 6 & 6 & 8 & 10 & 12 \\
$U_{p,l}$ mode  & 2, 5 & 4, 1 & 1, 7 & 4,-1 & 0, 9 & 3,-3 & 2,-5 & 1,-7 & 0,-9 \\
\hline
\hline
\end{tabular}
\end{center}
\caption{The azimuthal index ($m$) of the Zernike polynomial required to achieve significant coupling from an LG$_{33}$ incident beam into the other order 9 modes.}
\label{table:m_for_order_9}
\end{table}

Higher order Zernike polynomials represent higher order spatial frequencies, which generally have smaller amplitudes in the mirror surfaces.  Therefore we would expect the coupling caused by higher order polynomials, such as into LG$_{1-7}$ and LG$_{0-9}$, to be smaller than those caused by lower order polynomials.
We would also expect the polynomials with $m=2$, 4 and 6 to have a large effect on the beam
purity as they each couple from LG$_{33}$ into two other order 9 modes.  

\begin{figure}[t]
   \centering
   \includegraphics[scale=0.53]{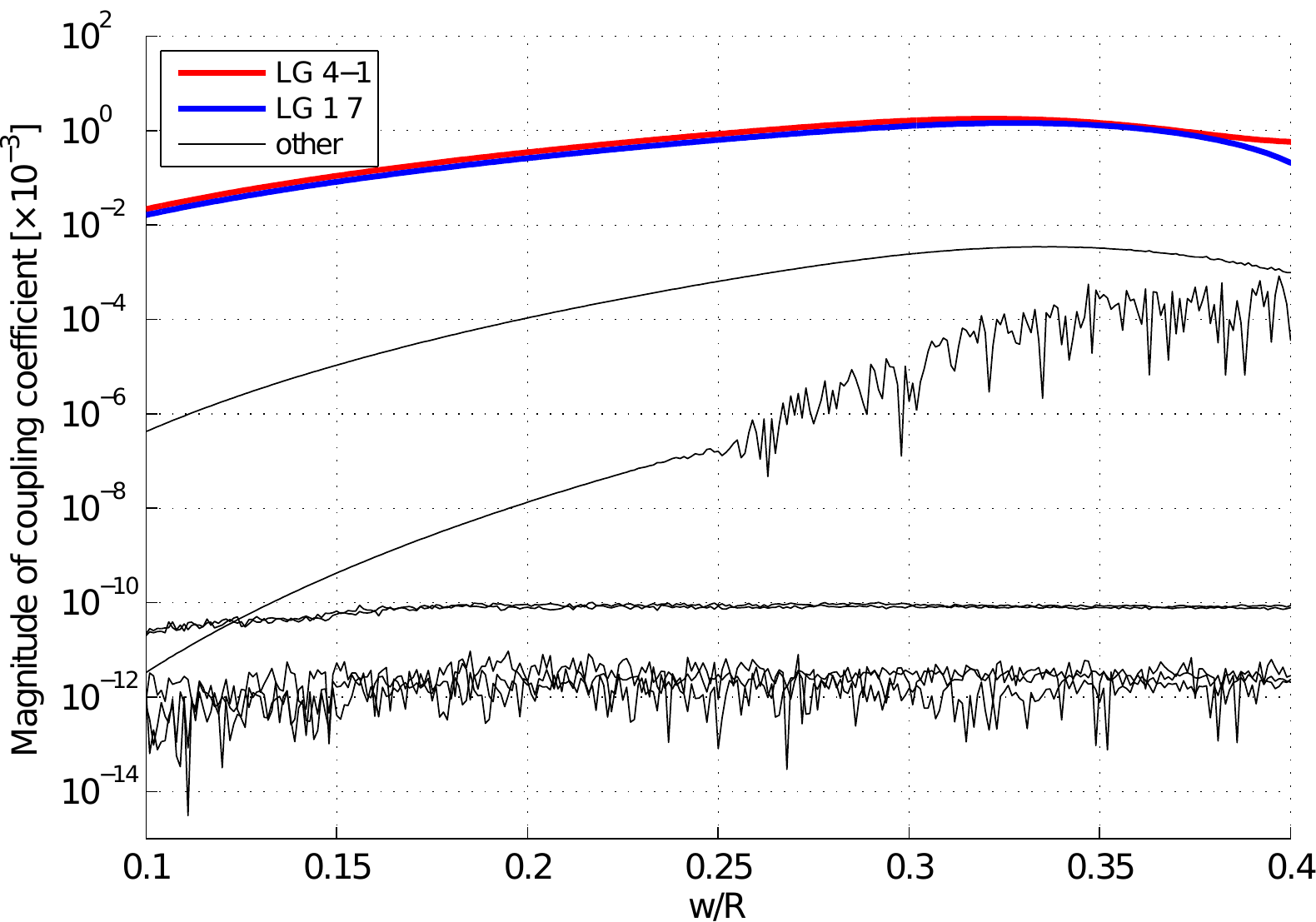}
   \caption{Plots of the coupling coefficients for different order 9 LG modes when an LG$_{33}$ beam is incident on a surface described by the Zernike polynomial Z$_{4}^{4}$.  The amplitude of the coefficients is plotted against the ratio of the beam radius, $w$, and the Zernike radius, $R$.  These plots show numerical results for the coefficients without any approximation.  The coefficients for LG$_{4-1}$ and LG$_{17}$ are significantly larger than those of the other modes.}
   \label{fig:numerical_coupling}
\end{figure}

Using Matlab the original integration (Eq.~\ref{eq:original_eq}) was performed numerically,
computing the coupling occurring from a mirror surface defined
completely by a single Zernike polynomial.
This particular example shows the results for Z$_{4}^{4}$, in which
we expect a large coupling from LG$_{33}$ into LG$_{17}$ and LG$_{4-1}$
(table~\ref{table:m_for_order_9}) and much less coupling into 
other order 9 modes.
The coupling coefficients between the LG$_{33}$ beam and all the 
other order 9 modes were calculated.  The numerical integration was 
carried out for a range of $\frac{w}{R}$ and the results, for a polynomial 
amplitude of 1 nm, are summarised in Fig.~\ref{fig:numerical_coupling}. 
In this plot the two largest coupling coefficients over this range of $\frac{w}{R}$ correspond to LG$_{4-1}$ and LG$_{17}$.  
The coefficients for the other LG modes are significantly smaller.  These results agree with our predictions.

\begin{figure}[t]
   \centering
   \includegraphics[scale=0.53]{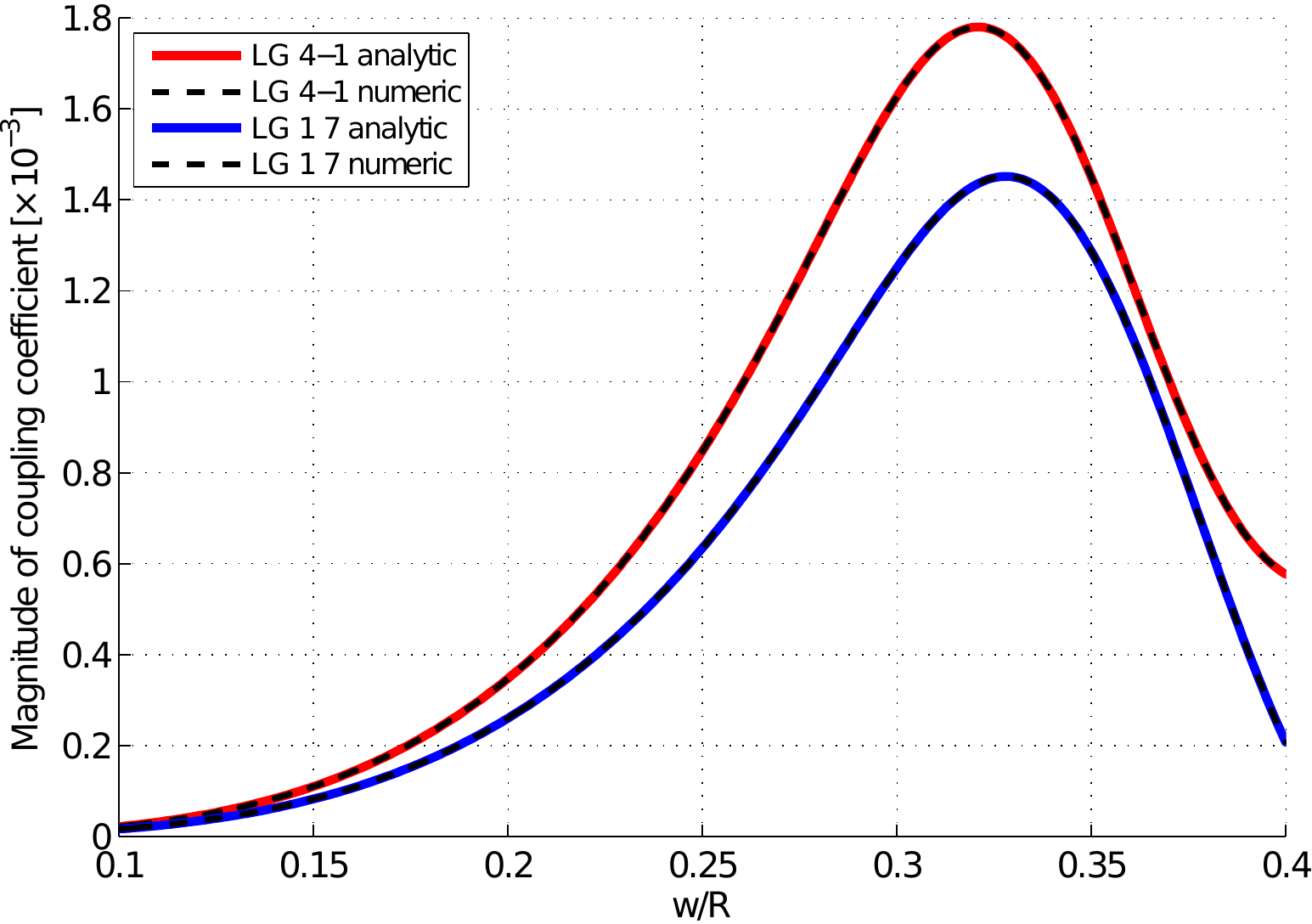} 
   \caption{A plot showing the coupling into the LG$_{4-1}$ and LG$_{17}$ modes from an LG$_{33}$ beam incident on a surface described by the Zernike polynomial Z$_{4}^{4}$.  The results are plotted against the relative beam radius on the mirror, $\frac{w}{R}$.  The results from an analytical approximation and a numerical integration method are plotted.}
   \label{fig:numerical_vs_analytical}
\end{figure}

Fig.~\ref{fig:numerical_vs_analytical} shows a comparison of the analytical results from
Eq.~\ref{eq:k_eq} with the numerical results. 
Here only the coupling coefficients for LG$_{4-1}$ and LG$_{17}$ are plotted as the 
analytical approach gives 0 for $m\neq |l-l'|$.  Over this range the two sets of numbers match up very well
and we consider this a good confirmation of the analytical approximation.

\subsection{Beam size and Zernike order}

Figs.~\ref{fig:numerical_coupling} and~\ref{fig:numerical_vs_analytical} seem to suggest that the coupling coefficients also depend on 
the beam size relative to the radius of the Zernike polynomial, or mirror radius.  However, this
ratio is typically not a free parameter. 
For a good reduction in thermal noise a large beam radius is desirable.  An upper limit for the beam width 
can be derived from optical loss due to beam clipping.
This so-called \emph{clipping loss} refers to the power lost over the mirror edges given 
by~\cite{Chelkowski09}:
\begin{equation}
l_{\mbox{clip}}  = 1-\int_{S} \left|U_{p,l}\right|^{2},
\end{equation}
where the integral represents the normalised power reflected by a perfect mirror of finite size (see Appendix~\ref{sec:acliploss}). 
In gravitational wave detectors the clipping loss should be lower than 100\,ppm (parts per million) and often an arbitrary requirement of 1\,ppm is used
during the interferometer design phase. This yields an optimal beam radius of $0.232\,R$  ($R$ being the mirror radius)
for an LG$_{33}$ beam.
A clipping loss of 100\,ppm instead leads to an optimal beam radius of $0.255\,R$.

\section{Cavity simulations with Advanced LIGO mirror maps}

\begin{figure}[t]
   \centering
   \includegraphics[scale=0.8]{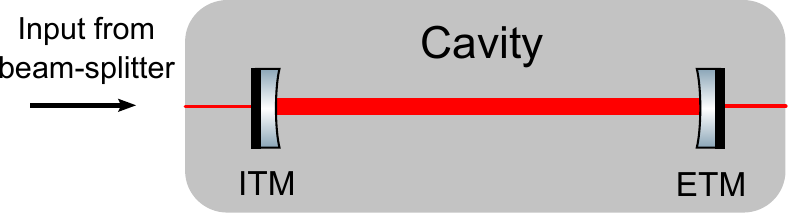} 
   \caption{The optical layout of an Advanced LIGO arm cavity.  Light from the beam-splitter is incident on the flat anti-reflective surface of the ITM.  The light circulates between the two highly reflective, curved surfaces of the ITM and ETM.  Light is transmitted by the cavity through the ETM.}
   \label{fig:ligo_cav}
\end{figure}

We want to investigate how the degeneracy of the order 9 modes affects the purity of an LG$_{33}$ mode in 
high finesse cavities.  The aim of this investigation is to assess the effects of higher-order mode degeneracy 
and derive requirements for the mirror surfaces which would result in an acceptably high LG$_{33}$ beam purity.  The Advanced LIGO cavities consist of two curved mirrors; the input test mass (ITM) and the end test mass (ETM) separated by an arm length of approximately 4\,km (Fig.~\ref{fig:ligo_cav}).  The design properties of these mirrors are summarised in table~\ref{table:mirror_props}~\cite{ITM, ETM}, giving a high finesse of 450.

\label{sec:ligo_cavs}

\begin{table}[t]
\begin{center}
\begin{tabular}{ccc}
\hline
\hline
Mirror & ITM & ETM \\
\hline
$R_{a}$ & 20\,ppm & 500\,ppm \\
$L_{a}$ & 1\,ppm & 1\,ppm \\
$T_{h}$ & 0.014 & 5\,ppm \\
$L_{h}$ & 0.3\,ppm & 0.3\,ppm \\
\hline
\hline
\end{tabular}
\end{center}
\caption{Optical properties of the mirrors designed for the Advanced LIGO arm cavities.  The two mirrors have a design thickness of 200 mm and index of refraction of 1.45.  $R$, $T$ and $L$ refer to the reflectivity, transmission and loss of power at the mirror.  $a$ and $h$ refer to the anti-reflective and highly reflective coated surfaces of each mirror.}
\label{table:mirror_props}
\end{table}

To simulate mirror surface distortions we used mirror maps measured from 
uncoated mirror substrates produced for Advanced LIGO \footnote{At the 
time of our analysis the coated mirrors were not yet available. The 
coating process can add further surface distortions so that some of the
results presented here might not be representative for the final 
Advanced LIGO cavities.}.
However, the Advanced LIGO cavities were not designed to be compatible 
with the LG$_{33}$ mode.
The LG$_{33}$ mode is more spatially extended than the LG$_{00}$ mode, 
and so experiences a larger clipping loss at the mirrors for a given 
beam spot size value.
For the Advanced LIGO cavity parameters, the clipping loss for the 
LG$_{33}$ mode is much larger than acceptable, at around 35\%.
It was therefore necessary to adjust the cavity length to bring the 
LG$_{33}$ clipping to a level that allowed
us to carry out a meaningful investigation.
Using a cavity length of 2802.9\,m we achieve similar clipping losses to 
those
experienced by LG$_{00}$ in the original cavities, for the 30\,cm aperture
represented by the mirror maps.
The results from this optical setup should be representative of longer 
cavities
with larger mirrors.
The simulated cavity parameters are summarised
in table~\ref{table:cavity_params}.

\begin{table}[b]
\begin{center}
\begin{tabular}{lccc}
\hline
\hline
Parameter & ITM R$_{c}$ & ETM R$_{c}$ & Cavity length \\
Value [m] & -1934 & 2245 & 2802.9 \\
\hline
\hline
\end{tabular}
\end{center}
\caption{Cavity parameters for simulations of Advanced LIGO style arm cavities~\cite{AdLIGO}.  The length of the cavity was reduced from the original length of 3994.5 m to prevent a large clipping loss when using LG$_{33}$ beams.}
\label{table:cavity_params}
\end{table}

\subsection{Laguerre-Gauss mode purity with Advanced LIGO mirror maps}
\label{sec:ad_ligo_maps}

For the purposes of this investigation the mirror map 
corresponding to the Advanced LIGO end test mass ETM08 was used; see Fig.~\ref{fig:ETM08}.  
To predict the direct couplings from this 
mirror we look at the Zernike polynomials representing the 
mirror surface.  This is achieved by performing a convolution between
the surface defined by the mirror map and the different Zernike polynomials:
\begin{equation}
\int_{S}Z_{map} \cdot Z_{n}^{m}=A_{n}^{m}\int_{S} Z_{n}^{m} \cdot Z_{n}^{m}
\end{equation}
where $Z_{map}$ is the surface defined by the mirror map and $A_{n}^{m}$ is the
amplitude of the corresponding Zernike polynomial in the surface.
The convolution was performed for all Zernike polynomials with $n \leq 30$.
The polynomials which cause significant coupling into the other order 9 modes ($m=2$, 4,$\dots$12) are summarised in table~\ref{table:ETM08_zernikes}.  Here the polynomials are ranked in order of the
power they couple into the other order 9 modes when an LG$_{33}$ beam is reflected from a surface described by the polynomial. 
\begin{table}[b]
\begin{center}
\begin{tabular}{lcccccc}
\hline
\hline
Z$_{n}^{m}$ polynomial & 2, 2 & 4, 2 & 4, 4 & 6, 2 & 10, 8 & other \\
$A_{n}^{m}$ [nm] & 0.908 & 0.202 & 0.213 & 0.124 & 0.116 & - \\
Power [ppm] & 4.66 & 0.331 & 0.0431 & 0.0099 & 0.0059 & $<$ 0.005 \\
\hline
\hline
\end{tabular}
\end{center}
\caption{Zernike polynomials present in the Advanced LIGO mirror map ETM08 which cause significant coupling from LG$_{33}$ into the other order 9 LG modes ($m = 2$, 4,$\dots$12).  The power coupled from LG$_{33}$ into the other order 9 modes by reflection from surfaces described by the individual polynomials is included, calculated from a coupling approximation.}
\label{table:ETM08_zernikes}
\end{table}
From this we can suggest which order 9 LG modes will have significant amplitudes in the simulated cavity.
The two polynomials which cause the largest individual power 
couplings have $m=2$.  The astigmatism (Z$_{2}^{2}$) in particular 
extracts a large amount of power from LG$_{33}$. 
Therefore, we would expect the LG$_{41}$ and LG$_{25}$ modes to have relatively large amplitudes in the cavity as the $m=2$ polynomials cause significant coupling into these modes.     

\begin{figure}[htbp]
   \centering
   \includegraphics[scale=0.5]{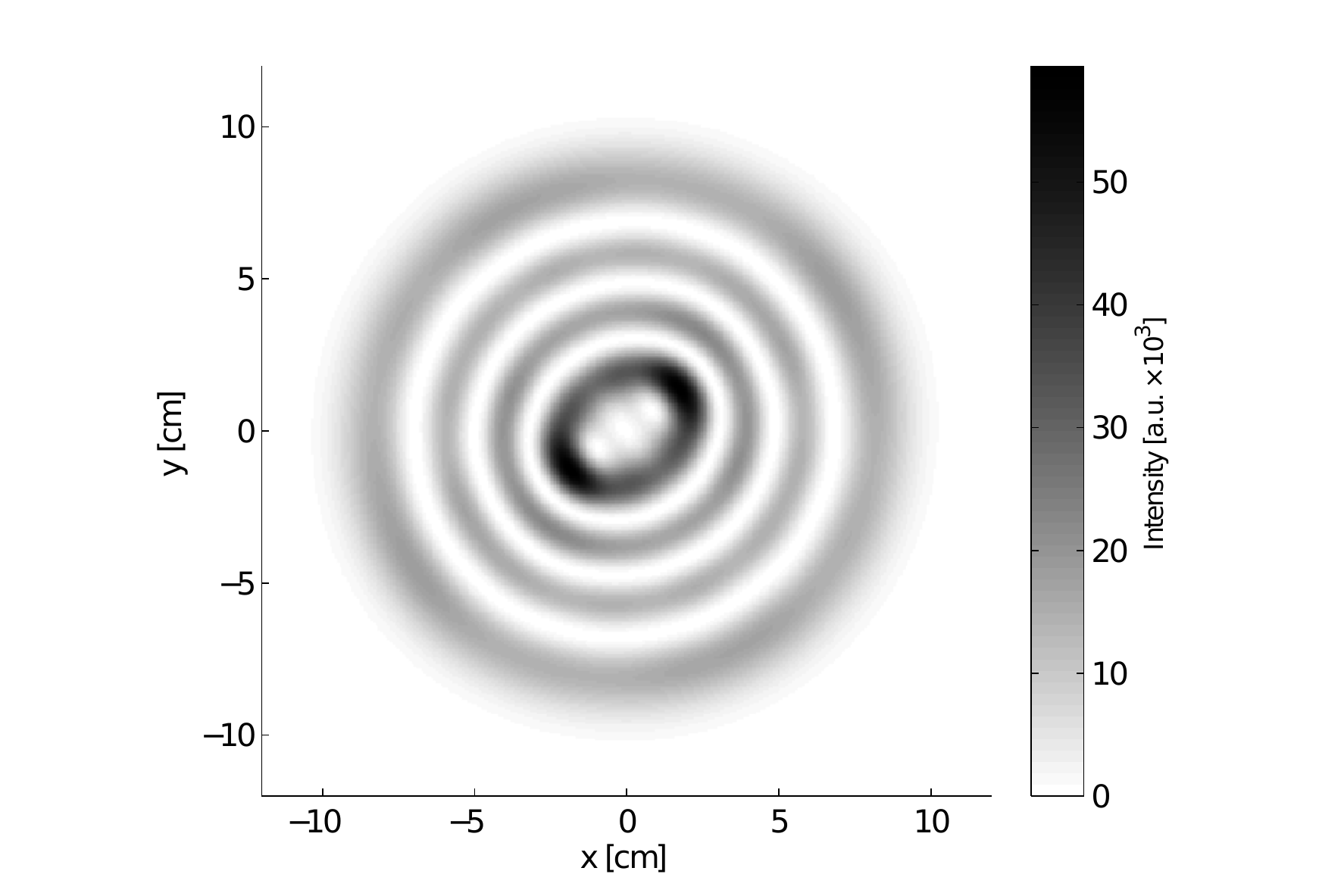} 
   \caption{A plot of the field circulating in a simulated high finesse cavity.  In the simulation the Advanced LIGO mirror map ETM08 is applied to the end test mass and a pure LG$_{33}$ beam is injected into the cavity.}
   \label{fig:ETM08_sim}
\end{figure}

The cavity defined in Sec.~\ref{sec:ligo_cavs} was simulated using the interferometer simulation tool \textsc{Finesse} \footnote{\textsc{Finesse} has been tested against an FFT propagation simulation, with higher order LG modes and mirror surface distortions.  The results suggest \textsc{Finesse} is suitable for this investigation~\cite{Bond}.}~\cite{Finesse, Finesse2}.  An input beam of pure LG$_{33}$ was used with the ETM08 mirror map applied to the end mirror and a perfect input mirror.  The cavity was tuned to be on resonance for the LG$_{33}$ mode and the beam circulating in the cavity was detected.  A plot of the circulating field is shown in Fig.~\ref{fig:ETM08_sim}.

The purity of an LG mode,  $U_{p,l}$, in a given beam $U$ is defined as $|c_{p,l}|^2$ where~\cite{Chu}:
\begin{equation}
c_{p,l}=\int_{S}U \ U_{p,l}^{*}
\end{equation}
The plot in Fig.~\ref{fig:ETM08_sim} (compared with the plot of LG$_{33}$ in Fig.~\ref{fig:LG9}) suggests that the 
circulating beam now contains modes other than LG$_{33}$.  The purity 
of the LG$_{33}$ beam in this simulated cavity was found to be 88.6\%.  
The power in the different modes present in 
the circulating field are summarised in table~\ref{table:ETM08_modes}.
\begin{table}[b]
\begin{center}
\begin{tabular}{lcccccccccc}
\hline
\hline
$U_{p,l}$ mode & 3, 3 & 4, 1 & 2, 5 & 4,-1 & 1, 7 &  other\\
$m_{SC}$ & - &  2  & 2 & 4 & 4 & - \\
Power (\%) & 88.6 & 5.70 & 5.02 & 0.333 & 0.313 & $<$ 0.05 \\
\hline
\hline
\end{tabular}
\end{center}
\caption{The power in the LG modes circulating in a cavity simulated with \textsc{Finesse} with an LG$_{33}$ input beam and the Advanced LIGO mirror map ETM08.  $m_{SC}$ refers to the azimuthal index of the Zernike polynomial required to cause significant coupling from LG$_{33}$ to the given mode.}
\label{table:ETM08_modes}
\end{table}  
The results of the decomposition show that the surface distortions of 
the end mirror cause significant
coupling into other LG modes, particularly the other order 9 modes.
The distortions also cause coupling into modes of other orders, but 
these are not resonant at the same cavity tuning as the LG$_{33}$ mode 
and so are strongly suppressed.

Other than LG$_{33}$ the
two largest modes in the cavity are LG$_{41}$ (5.7 \%) and
LG$_{25}$ (5.0 \%). This agrees with the predictions made by
studying the Zernike content of the ETM08 mirror map (table
\ref{table:ETM08_zernikes}).
LG$_{4-1}$ and LG$_{17}$ also have relatively large amplitudes
in the cavity. This may be due to the direct coupling out of the 
LG$_{33}$ caused by the Z$_{4}^{4}$ polynomial.
However, the LG$_{4-1}$ and LG$_{17}$ modes can also be strongly coupled 
out of the LG$_{41}$ and
LG$_{25}$ modes via $m=2$ polynomials, further contributing to the
effect these polynomials have on the purity of the beam.
Overall the coupling process in a cavity
is complicated by these multiple cross-couplings, but the results of 
this simulation suggest that the
direct coupling from a mirror surface is the dominant effect on the
mode content of the circulating beam. A theoretical understanding of the 
direct coupling has therefore allowed us to make valid predictions about 
the resulting mode content.

\subsection{Frequency splitting}

The presence of mirror surface distortions not only causes coupling between LG modes
but introduces additional phase shifts of the modes.
This results in slight shifts of the resonance frequency of individual modes.  
These shifts in resonance frequency depend on the particular mode 
and so modes of the same order will be resonant at slightly different frequencies.
Thus the mode degeneracy can be broken.  We will refer to 
this effect as \emph{frequency splitting}.  
For the frequency splitting to be effective the shifts in resonance frequency 
must be larger than the
cavity bandwidth in order to separate the resonance peaks of the 
different order 9 modes.

Using the ETM08 mirror map the high finesse cavity  defined in Sec.~\ref{sec:ligo_cavs} was simulated.  The beam circulating in the cavity was detected as the
laser frequency was tuned around the cavity resonance.
The maximum power of the order 9 modes in the cavity and the difference in their resonance frequencies is summarised in 
table~\ref{table:f_split}.  
The frequency splitting is of the order of 10\,Hz, smaller than the cavity bandwidth of 120\,Hz
and so is not sufficient to completely break the degeneracy.
The result is 10 \emph{quasi-degenerate} modes.
The resonance frequencies are slightly different and so when the 
cavity is tuned to the resonance of the LG$_{33}$ mode
the other modes will be slightly suppressed.
However, the frequency splitting is small and therefore
these modes will still have relatively large amplitudes in the cavity.
The coupling into order 9 modes is therefore still the dominant effect on the mode purity.     

\begin{table*}[t]
\begin{center}
\begin{tabular}{lcccccccccc}
\hline
\hline
$U_{p,l}$ mode & 3,  3 & 4,  1 & 2,  5 & 4, -1 & 1,   7 & 3, -3 & 0, -9 & 2, -5 & 0,  9 & 1,  -7 \\
Power [W] & 221.8 & 14.38 & 12.62 & 0.865 & 0.802 & 0.102 & 0.038 & 0.038 & 0.032 & 0.001 \\
Frequency [Hz] & 0 & 1.3 & 0.7  & 7.0 & 6.6  & 7.6 & -7.3 & -16.1 & -14.8 & 9.3 \\
\hline
\hline
\end{tabular}
\end{center}
\caption{A summary of the power in each order 9 mode in a simulated cavity where the mirror map ETM08 is applied to the end mirror and the input mode is LG$_{33}$.  The frequency is tuned around the cavity resonance and the difference in resonance frequency, compared to LG$_{33}$, is included for each mode.}
\label{table:f_split}
\end{table*}

\subsection{Laguerre-Gauss mode purity with improved mirror maps}

\begin{figure}[t]
   \centering
   \includegraphics[scale=0.5]{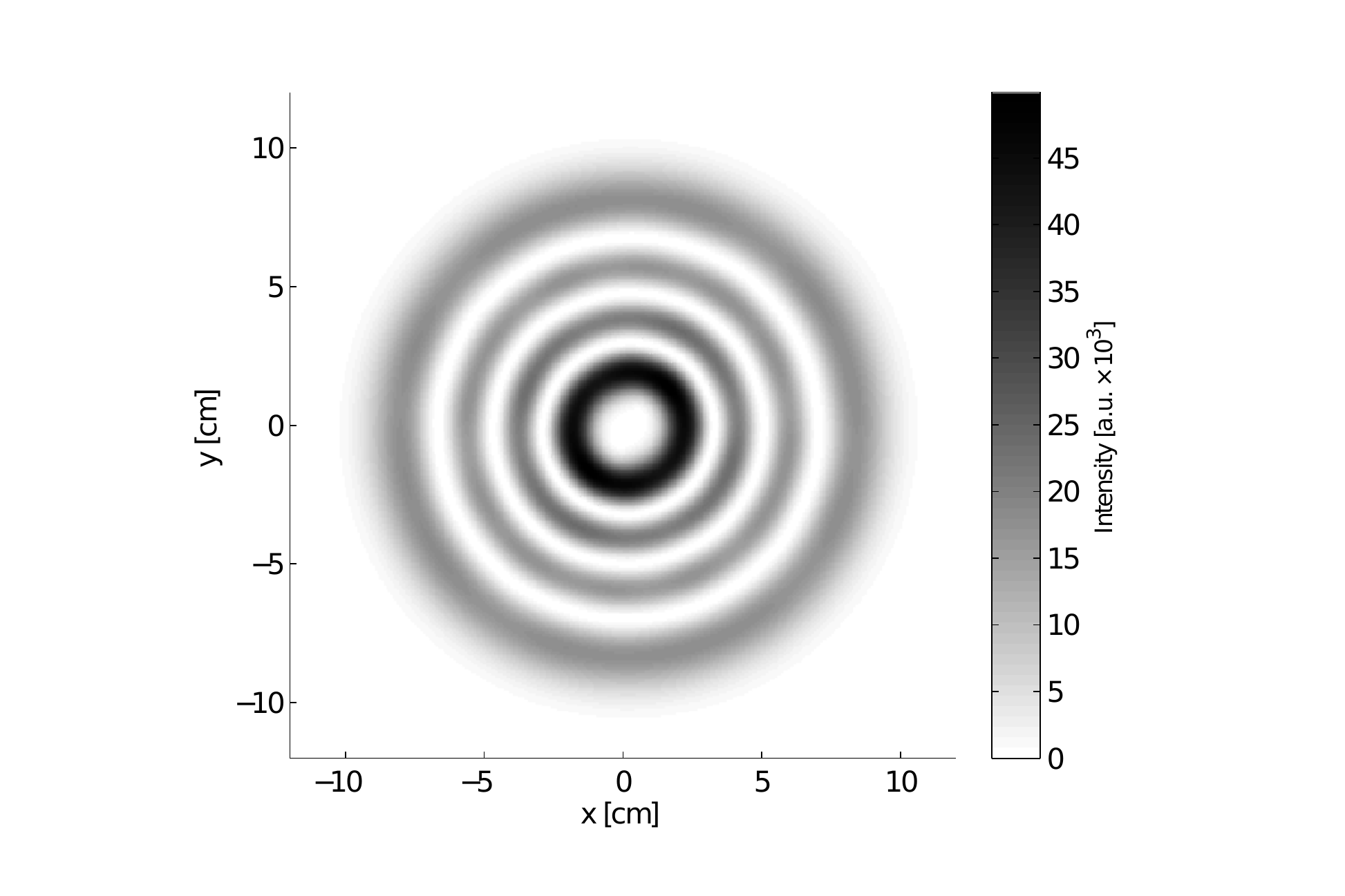} 
      \caption{A plot of the field circulating in a cavity simulated with an input mode of pure LG$_{33}$ and the Advanced LIGO ETM08 mirror map, with astigmatism removed.}
   \label{fig:ETM08-Z2s_sim}
\end{figure}

We want to find ways in which the mirror maps could be improved for the specific application of the LG$_{33}$ beam.  By considering the mirror surfaces analytically it appears that reducing the astigmatism for this particular map could improve the purity of LG$_{33}$ in our simulated cavity.  To demonstrate this the astigmatism was removed from the ETM08 mirror map.  The cavity was then simulated with this processed map, with the resulting circulating field detected and decomposed in to LG modes.  The circulating beam is plotted in Fig.~\ref{fig:ETM08-Z2s_sim}.  Simply comparing this plot with the original circulating beam (Fig.~\ref{fig:ETM08_sim}) suggests that the purity of the field has increased.  The LG content of the beam is summarised in table~\ref{table:ETM08-Z2s_modes}.  The purity of the circulating LG$_{33}$ beam is now 99.5\%, a significant improvement from the original results. 

\begin{table}[b]
\begin{center}
\begin{tabular}{lccccccc}
\hline
\hline
$U_{p,l}$ mode & 3, 3 & 4, 1 & 2, 5 & 1, 7 & 4,-1 & 0,-9 & other \\
$m_{SC}$ & - & 2 & 2 & 4 & 4 & 12 & - \\
Power [\%] & 99.5 & 0.231 & 0.208 & 0.0524 & 0.0165 & 0.0137 & $<$ 0.01 \\
\hline
\hline
\end{tabular}
\end{center}
\caption{The power in the LG modes circulating in a simulated cavity.  The cavity was simulated with an LG$_{33}$ input beam and with the ETM08 mirror map with astigmatism removed.  $m_{SC}$ refers to the azimuthal index of the Zernike polynomial required to cause significant coupling from LG$_{33}$ to the given mode.}
\label{table:ETM08-Z2s_modes}
\end{table}

The results of the decomposition show that the power in both the LG$_{41}$ and LG$_{25}$ modes has decreased significantly, as predicted.  The power in the other modes has also noticeably decreased.  This result suggests that the astigmatism is a major factor in coupling from the LG$_{33}$ mode, not only for its direct coupling into LG$_{25}$ and LG$_{41}$ but for the coupling from these new modes into other modes of order 9.  For this setup we can conclude that the astigmatism should be limited in the mirror surfaces to reduce the problems caused by higher order mode degeneracy.

\subsection{Mirror requirements for LG$_{33}$}

In order to use LG$_{33}$ in GW detectors we require certain Zernike polynomials in the mirrors 
to be smaller than in the current state of the art mirrors, in order to achieve an acceptable 
beam purity in the cavity.  
Here we investigate the direct coupling from ETM08 and suggest 
limits to the amplitudes of specific polynomials in the mirror surfaces, presenting an
Advanced LIGO mirror map adapted for the use of LG$_{33}$.

\begin{figure}[b]
   \centering
   \includegraphics[scale=0.55]{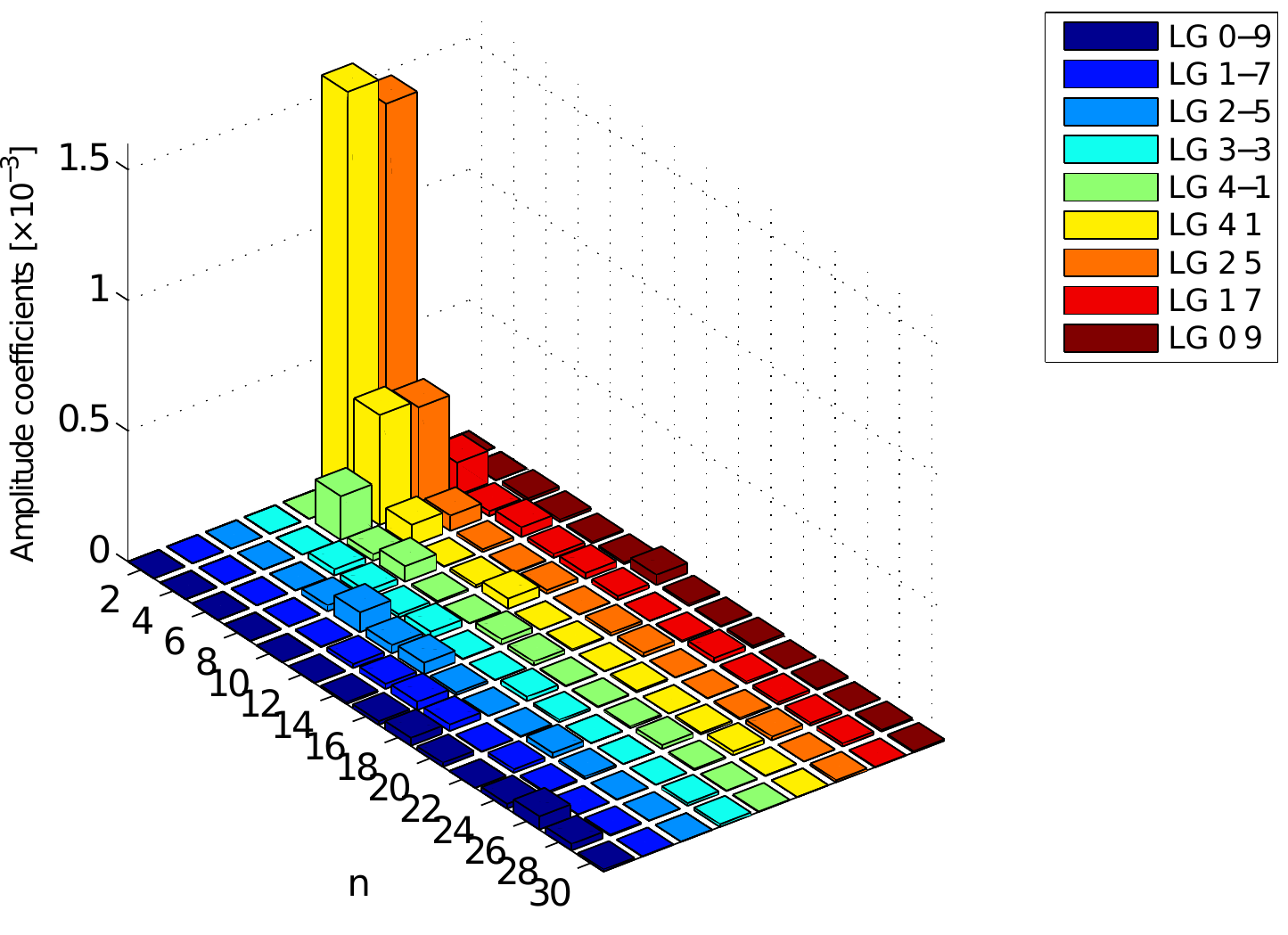} 
   \caption{A bar chart showing the coupling, $k_{p,l,p',l'}^{n,m}$, into the order 9 modes when an LG$_{33}$ beam is incident on a surface described by the Zernike polynomial with $m$ required to cause significant coupling.  The Zernike amplitudes correspond to those in the ETM08 mirror map.}
   \label{fig:etm08_cc}
\end{figure}

Using our theoretical analysis we can identify the particular Zernike polynomials to 
reduce in the mirrors.  We have 
already seen that the Zernike polynomials with odd values of 
$n$ and with $m > 12$ don't have a large effect on the purity. 
To assess the other polynomials we use Eq.~\ref{eq:k_eq} to approximate the 
coupling into order 9 modes caused by the polynomials in the ETM08 mirror map,
for the optical setup defined in Sec.~\ref{sec:ligo_cavs}.
For each order 9 LG mode the coupling was calculated for the Zernike polynomials with $n=2$, 4 $\dots$30 and with $m$ required to give significant coupling.
Fig.~\ref{fig:etm08_cc} represents these coupling coefficients.       
This chart shows that the largest couplings occur from Z$_{2}^{2}$, 
into LG$_{25}$ and LG$_{41}$, as previously suggested.  There 
is also some strong coupling from Z$_{4}^{2}$ and Z$_{4}^{4}$.  
The other couplings are significantly smaller.  Therefore, the first step in 
modifying the mirrors for LG$_{33}$ is to limit these 3 
polynomials to give similar couplings to the higher order polynomials.   

\begin{figure}[t]
   \centering
   \includegraphics[scale=0.5]{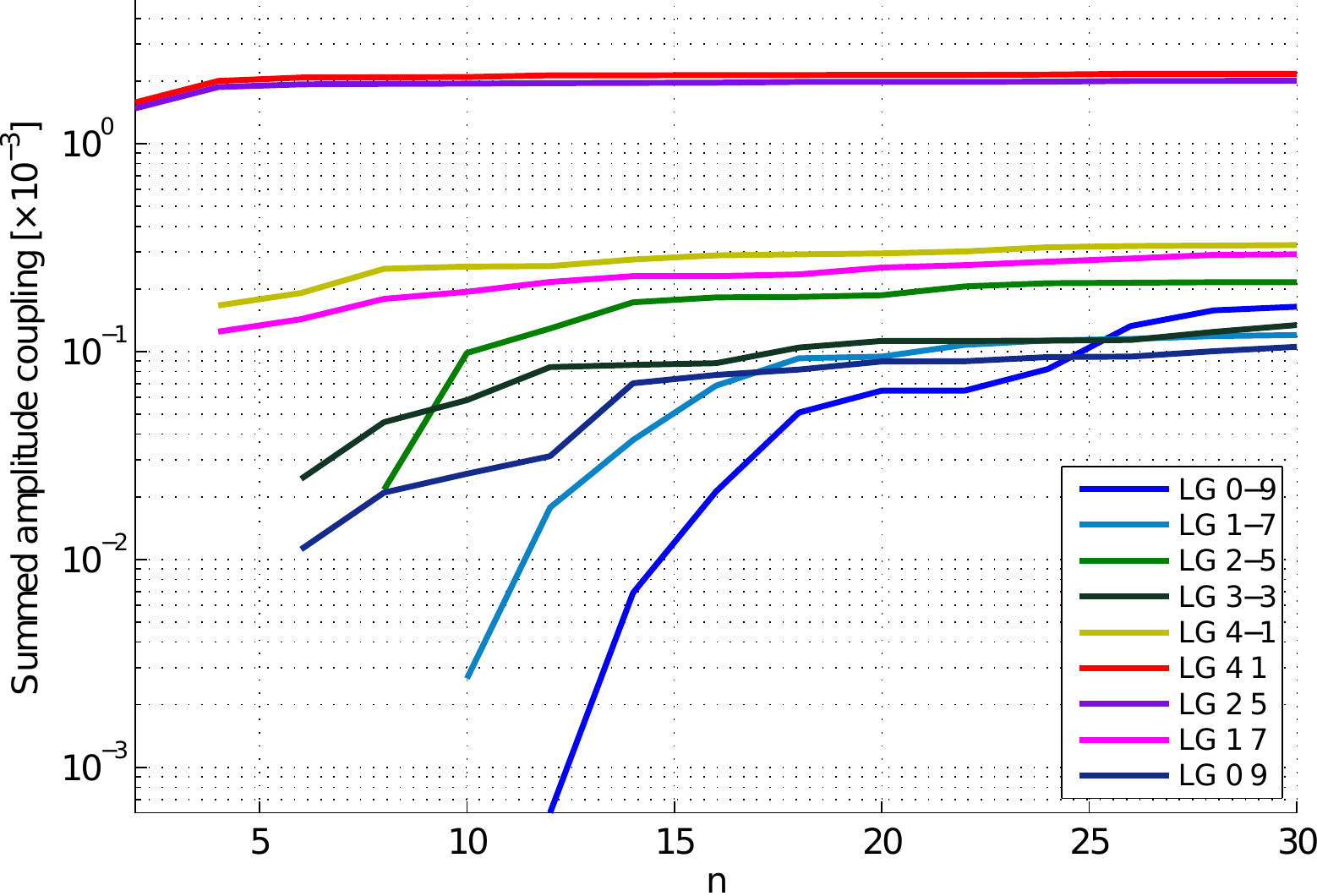} 
   \caption{Plots showing an approximation of the coupling from LG$_{33}$ into the other order 9 LG modes from the ETM08 mirror map as higher order ($n$) Zernike polynomials are included in our model.  The sum of the coupling from polynomials with order $\leq n$ are plotted for each mode.}
   \label{fig:etm08_sum_cc}
\end{figure}

Another plot illustrating the coupling from this map is shown in 
Fig.~\ref{fig:etm08_sum_cc}.  In this plot the maximum possible coupling from the ETM08 
mirror map is estimated using our analytical approximation.  For each 
order 9 mode the sum of the coupling is calculated for all Zernike polynomial
orders smaller than $n$.
The plot illustrates the magnitude of the direct coupling expected as we 
include higher order Zernike polynomials in our model. 
Fig.~\ref{fig:etm08_sum_cc} can be used to illustrate how this 
particular map, ETM08, should be adapted for LG$_{33}$.  The 
coupling into LG$_{25}$ and LG$_{41}$ is around 10 times 
greater than any other coupling.  The coupling into these two 
modes is also not significantly increased by including the higher 
order modes.  Therefore, limiting the lower order polynomials 
with $m=2$ will greatly reduce the coupling into these 
two modes and the overall coupling into order 9 modes.  
From this plot we conclude that the overall coupling into order 9 
modes can be reduced by around a factor of 10 by reducing the 
lower order polynomials.  Reducing the coupling
further will involve limits on multiple polynomials.

By assessing the direct coupling from ETM08 we can set requirments 
for the lower order Zernike polynomials.  When an LG$_{33}$ 
beam is incident on the ETM08 mirror map in our setup the reflected 
beam is predominantly LG$_{33}$.  However, we find 31\,ppm (parts per million)
of the power is now in other modes, with 6.8\,ppm in the other 
order 9 modes.  Table~\ref{table:ETM08_zernikes} shows the 
power coupled from surfaces described by the polynomials 
present in the ETM08 map, from LG$_{33}$ into the other order 9 modes.  For this map  
we consider the polynomials causing a large amount of coupling 
as those coupling more than 0.01\,ppm into the order 9 
modes; Z$_{2}^{2}$, Z$_{4}^{2}$ and Z$_{4}^{4}$.  
For LG$_{33}$ we require these polynomials to be limited in the 
mirror surfaces.  The requirements for these polynomials
were calculated to give power couplings of 0.01\,ppm into the 
other order 9 modes and are 
summarised in table~\ref{table:limits}.
\begin{table}[t]
\begin{center}
\begin{tabular}{l c c c}
\hline
\hline
 $Z_{n}^{m}$ polynomial & 2, 2 & 4, 2 & 4, 4  \\
Amplitude [nm] & 0.042 & 0.035 & 0.100 \\ 
\hline
\hline
\end{tabular}
\end{center}
\caption{A summary of the amplitude requirements for the Zernike polynomials required to give individual couplings of 0.01\,ppm from LG$_{33}$ into the other order 9 modes in the ETM08 mirror map.}
\label{table:limits}
\end{table}
These amplitude limits were applied to the ETM08 mirror map, resulting in coupling of 
19\,ppm into modes other than LG$_{33}$ and
0.043\,ppm into the other order 9 modes.  The cavity defined in Sec.~\ref{sec:ligo_cavs} was simulated 
with this limited map, resulting in 815\,ppm impurity in the circulating beam.  
This is a very good improvement on the original impurity of 0.114, illustrating that 
a high beam purity is achievable with these mirror requirements.  To achieve an even higher
beam purity will involve reducing the amplitudes of these polynomials further, as well as 
additional Zernike requirements.

\section{Conclusion}
We have investigated the coupling which occurs when Laguerre-Gauss modes are incident on a mirror with surface distortions.  Taking an analytical approach we used Zernike polynomials to represent mirror surface distortions and derived an approximate equation for the significant coupling when an LG mode is reflected from a surface defined by a particular Zernike polynomial.  This derivation resulted in a condition for significant coupling, $m=|l-l'|$, where $m$ is the azimuthal index of the Zernike polynomial and $l$ and $l'$ are the azimuthal indices of the incident and coupled modes respectively.  This is a significant result as it allows us to predict which order 9 modes will be largely coupled by particular Zernike polynomials and suggest which modes will have large amplitudes in the arm cavities

We investigated the performance of LG$_{33}$ in high finesse cavities by simulation with Advanced LIGO mirror maps.  This illustrated the degraded purity of the circulating beam in realistic cavities due to higher order mode degeneracy.  The results were then analysed by looking at the Zernike polynomials representing our example mirror map.  The analysis and results were consistent with the predictions made from Eq.~\ref{eq:k_eq}.  This suggested that astigmatism was causing a significant amount of coupling, particularly into the LG$_{41}$ and LG$_{25}$ modes.  This was confirmed when the cavity was simulated again with the astigmatism removed from the mirror map and we observed a dramatic increase in the LG$_{33}$ mode purity.

The analytical description enabled us to identify the specific Zernike polynomials which cause large couplings as well as the LG modes which would dominate as a result.  Using this we were able to derive certain requirements for our example mirror map, ETM08, in terms of limits on the amplitudes of the Zernike polynomials Z$_{2}^{2}$, Z$_{4}^{2}$ and Z$_{4}^{4}$ (table~\ref{table:limits}).  Using this map the resulting circulating beam impurity was found to be 815\,ppm, a significant reduction from the original impurity of 0.114.

This investigation has demonstrated that a high beam purity is achievable using an LG$_{33}$ beam when modifications are made to the low order Zernike polynomials in Advanced LIGO mirrors.  Implementing the LG$_{33}$ beam in gravitational wave detectors will be challenging as we require very small amplitudes on these lower order polynomials.  We should also consider that the example mirror surfaces considered here refer to uncoated substrates.  The coating process is likely to add to the lower order features in the mirror surfaces.  However, using this analytical approach we can derive specific requirements for the mirror surfaces
leading to designs for suitable mirrors for these higher order beams.         

\section*{Acknowledgments}
We would like to thank GariLynn Billingsley for providing the Advanced LIGO mirror surface maps and for advice and support 
on using them. We would also like to thank David Shoemaker and Stefan Hild for useful discussions. 
This work has been supported by the Science and Technology Facilities Council and the European Commission (FP7 Grant Agreement 211743). This document has been assigned the LIGO Laboratory document number LIGO--P1100081.

\appendix
\section{Derivation of coupling coefficients}
\label{sec:aderivation}

The product of two Laguerre-Gauss modes is:
\small
\begin{equation}
    \begin{split}
       U_{p,l}  U^{*}_{p',l'} = {}&  \frac{1}{w^{2}}\frac{2}{\pi}\sqrt{\frac{p!p'!}{(|l|+p)!(|l'|+p')!}} \\
       {}&\times  \exp{\left(i\left(2p+|l|-2p'-|l'|\right)\Psi\right)} \\
          {}& \left(\frac{\sqrt{2}r}{w}\right)^{|l|+|l'|} L^{|l|}_{p}\left(\frac{2r^{2}}{w^{2}}\right) L^{|l'|}_{p'}\left(\frac{2r^{2}}{w^{2}}\right)  \\
          {}& \exp{\left(-\frac{2r^{2}}{w^{2}}\right)} \exp{\left(i\phi\left(l-l'\right)\right)}
    \end{split}
 \end{equation}
\normalsize
The following derivation follows from Eq.~\ref{eq:m_condition}.  Currently we are concerned with the magnitude of the 
coupling coefficients, so we ignore any constant phase shifts and integrate with respect to $\phi$:
\begin{equation}
   \begin{split}
     k^{n,m}_{p,l,p',l'}    = {}&  \left| \int_{0}^{R} \frac{2}{\pi w^{2}}\sqrt{\frac{p!p'!}{(|l|+p)!(|l'|+p')!}}  2kA_{n}^{m}\pi R_{n}^{m}(r) \right.    \\
     {}& \times \left(\frac{\sqrt{2}r}{w} \right)^{|l|+|l'|} L_{p}^{|l|} \left(\frac{2r^{2}}{w^{2}} \right) L_{p'}^{|l'|}\left(\frac{2r^{2}}{w^{2}} \right) \\
         {}&   \left. \times\exp{\left(-\frac{2r^{2}}{w^{2}} \right)} rdr \right| \\
   \end{split}
   \label{eq:k1}
\end{equation}
In order to further simplify the equation the following variable substitution is made:
\begin{equation}
x=\frac{2r^{2}}{w^{2}}
\end{equation}
and a new limit to the integral:
\begin{equation}
X=\frac{2R^{2}}{w^{2}}
\end{equation}
where $R$ is the Zernike radius.  This gives the integral:
\begin{equation}
   \begin{split}
     k^{n,m}_{p,l,p',l'}    = {}&  kA_{n}^{m}\sqrt{\frac{p!p'!}{(|l|+p)!(|l'|+p')!}}        \\
         {}&   \times \left| \int_{0}^{X} R_{n}^{m}(x) x^{\frac{1}{2}(|l|+|l'|)} \right. \\
         {}& \times \left. L_{p}^{|l|} (x) L_{p'}^{|l'|}(x) \exp{(-x)} dx \right| 
   \end{split}
\end{equation}
Substituting in the sums representing the Laguerre polynomials and the radial Zernike function as a function of $x$ gives:
\begin{equation}
   \begin{split}
      k_{p,l,p',l'}^{n,m} = {}& A_{n}^{m}k \sqrt{p!p'!(p+|l|)!(p'+|l'|)!} \\
         {}&  \times \left| \sum_{i=0}^{p} \sum_{j=0}^{p'} \sum_{h=0}^{\frac{1}{2}(n-m)}  \frac{(-1)^{i+j+h}} {(p-i)!(|l|+i)! i!}  \right. \\
         {}& \times \frac{1}{(p'-j)!(|l'|+j)!j!} \frac{(n-h)!}{(\frac{1}{2}(n+m)-h)!} \\
         {}& \left. \times  \frac{1}{ \left(\frac{1}{2}(n-m)-h\right)! h!} \frac{1}{X^{\frac{1}{2}(n-2h)}} \right. \\         
         {}& \left. \times \int_{0}^{X} x^{i+j-h+\frac{1}{2}(|l|+|l'|+n)} \exp{(-x)} dx \right| \\
   \end{split}
\end{equation}
This type of integral results in the lower incomplete gamma function:
\begin{equation}
\gamma(a,x)=\int_{0}^{x} t^{a-1}e^{-t}dt
\end{equation}
When $a$ is equal to $n$, an integer, the function is given by the following sum:
\begin{equation}
\gamma(n,x)=(n-1)! \left(1-e^{-x} \sum_{k=0}^{n-1} \frac{x^{k}}{k!} \right)
\end{equation}
Therefore, the final equation for this approximation of the magnitude of the coupling coefficients is given by:
\begin{equation}
   \begin{split}
            k_{p,l,p',l'}^{n,m} = {}& A_{n}^{m}k \sqrt{p!p'!(p+|l|)!(p'+|l'|)!} \\
         {}&  \times \left| \sum_{i=0}^{p} \sum_{j=0}^{p'} \sum_{h=0}^{\frac{1}{2}(n-m)}  \frac{(-1)^{i+j+h}} {(p-i)!(|l|+i)! i!}  \right. \\
         {}& \times \frac{1}{(p'-j)!(|l'|+j)!j!} \frac{(n-h)!}{(\frac{1}{2}(n+m)-h)!} \\
         {}& \left. \times  \frac{1}{ \left(\frac{1}{2}(n-m)-h\right)! h!} \frac{1}{X^{\frac{1}{2}(n-2h)}} \right. \\           {}& \left. \times \ \gamma (i+j-h+\frac{1}{2}(|l|+|l'|+n)+1,X) \ \right|  \\
   \end{split}
   \label{eq:ak_eq}
\end{equation}

\section{Clipping loss}
\label{sec:acliploss}
Clipping loss is given by:
\begin{equation}
l_{\mbox{clip}}  = 1-\int_{S} \left|U_{p,l}\right|^{2}
\end{equation}
where the integral represents the normalised power reflected by a perfect mirror.  $S$ defines an infinite plane perpendicular to the beam axis.  The magnitude squared of an LG mode is:
\begin{equation}
\begin{split}
\left|U_{p,l}\right|^{2}= {}& \frac{1}{w^{2}}  \frac{2p!}{\pi(|l|+p)!} \left( \frac{2r^{2}}{w^{2}} \right)^{|l|} \\
{}& \times \left( L^{|l|}_{p}\left(\frac{2r^{2}}{w^{2}} \right) \right)^{2} \exp{\left( -\frac{2r^{2}}{w^{2}} \right)} \\
\end{split}
\end{equation}
where $w$ is the beam radius, $p$ and $l$ are the mode indices and $r$ is the radial position.  Integrating over the surface and taking a similar approach as for the coupling coefficients we get:
\begin{equation}
\begin{split}
l_{\mbox{clip}}= {}&1- p!(p+|l|)! \sum_{m=0}^{p}\sum_{n=0}^{p} \frac{(-1)^{n+m}}{(p-n)!(p-m)!} \\
{} &\times \frac{1}{  (|l|+n)!(|l|+m)!n!m!}\ \gamma(|l|+n+m+1,X) \\
\end{split}
\end{equation}
where $X=\frac{2R^{2}}{w^{2}}$, $R$ is the radius of the mirror and $\gamma$ is the lower incomplete gamma function.

\section{Zernike composition of a mirror surface}
\label{sec:azernike}
\begin{figure}[h]
   \centering
   \includegraphics[scale=0.5]{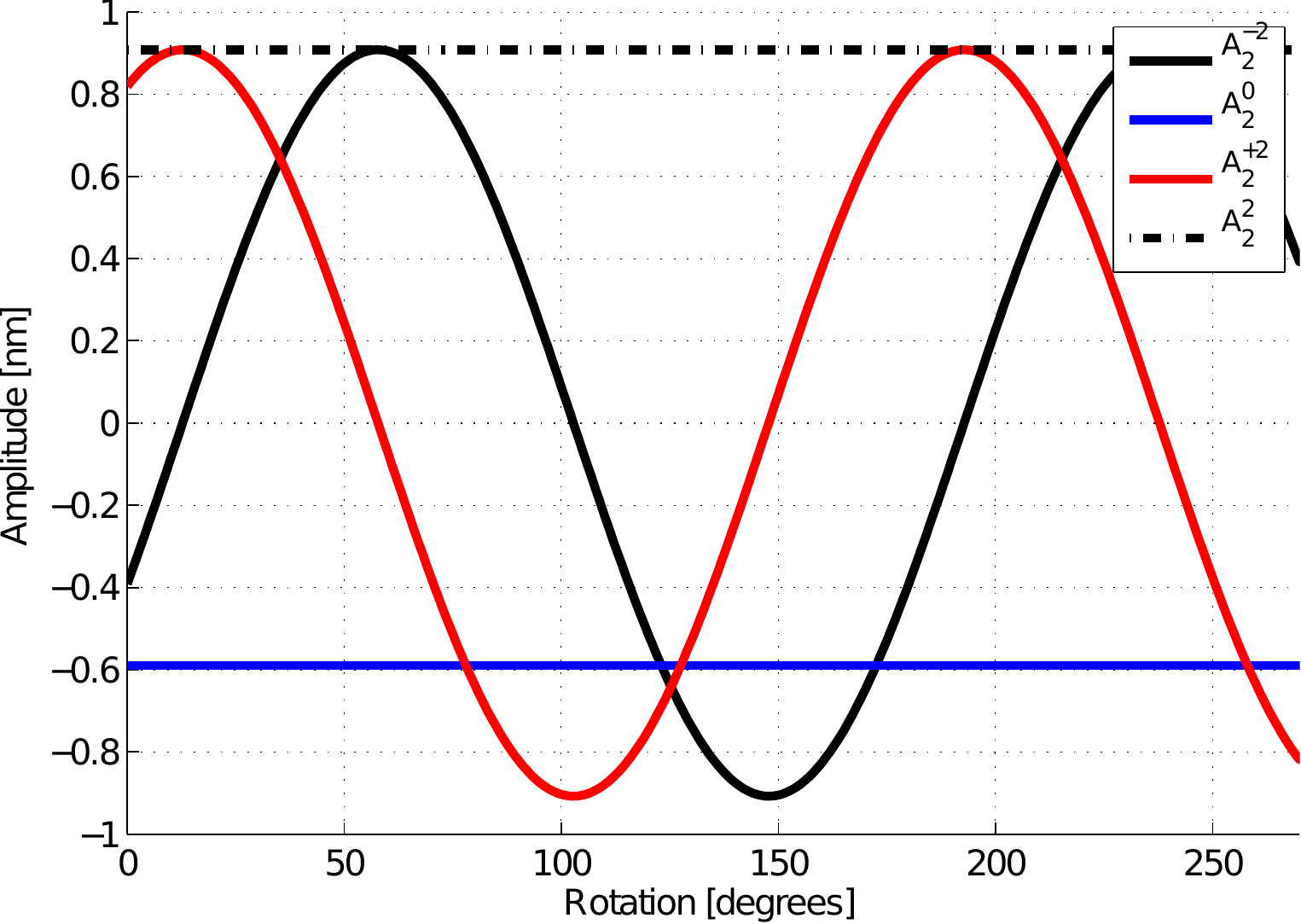} 
   \caption{Plots of the amplitudes of the order 2 Zernike polynomials present in the surface described by the ETM08 Advanced LIGO mirror map as a function of the rotation of the surface.}
   \label{fig:ETM08_zs_rotated}
\end{figure}

For certain Zernike polynomials (those with non-zero $m$) their amplitudes in a surface depend on the orientation of that surface with respect to the Zernike surface.  For example, consider the two polynomials responsible for astigmatism, Z$_{2}^{\pm2}$.  The two polynomials actually describe the same shape, with one just rotated by 90 degrees with respect to the other.  Therefore, rotating a surface, such as the one described by mirror map ETM08, will change the amplitudes of these two polynomials within the surface.  Fig.~\ref{fig:ETM08_zs_rotated} illustrates this effect.  The plots show the amplitudes of the order 2 Zernike polynomials present in the ETM08 mirror surface as it is rotated.  As expected the amplitude of the Z$_{2}^{0}$ polynomial remains constant as it has no angular dependence.  The amplitudes of the Z$_{2}^{\pm2}$ polynomials oscillate and, at a certain rotation (around 120$^{\circ}$) the astigmatism of the surface is completely described by Z$_{2}^{+2}$, and 90$^{\circ}$ later completely described by Z$_{2}^{-2}$.  The root mean squared amplitude ($A_{2}^{2}$) of the polynomials remains constant.

\section{Analysis of Zernike approach}

We have used Zernike polynomials to describe mirror surface distortions and analyse the coupling that occurs from LG$_{33}$ into other order 9 modes.  This approach appears to be very suitable as we have been able to identify specific polynomials which extract significant amounts of power from the input mode.  However, there is an alternative method used to investigate mirror surface distortions, which involves looking at the spatial frequencies present in real mirrors.  In this section we compare these two methods.

To look at the spatial frequencies present in realistic mirrors we perform a 2D Fourier transform of the surface height data.  The resulting spectra is then analysed and synthetic maps are created with the same spatial frequencies.  This method focuses on identifying particular spatial wavelengths which cause a large degree of coupling from LG$_{33}$.  Many synthetic maps are created and used in simulations of gravitational wave detectors.  A statistical approach is then taken to determine the extent of the coupling when specific spatial frequencies are present in the mirror.

\begin{figure}[htbp]
   \centering
   \includegraphics[scale=0.55]{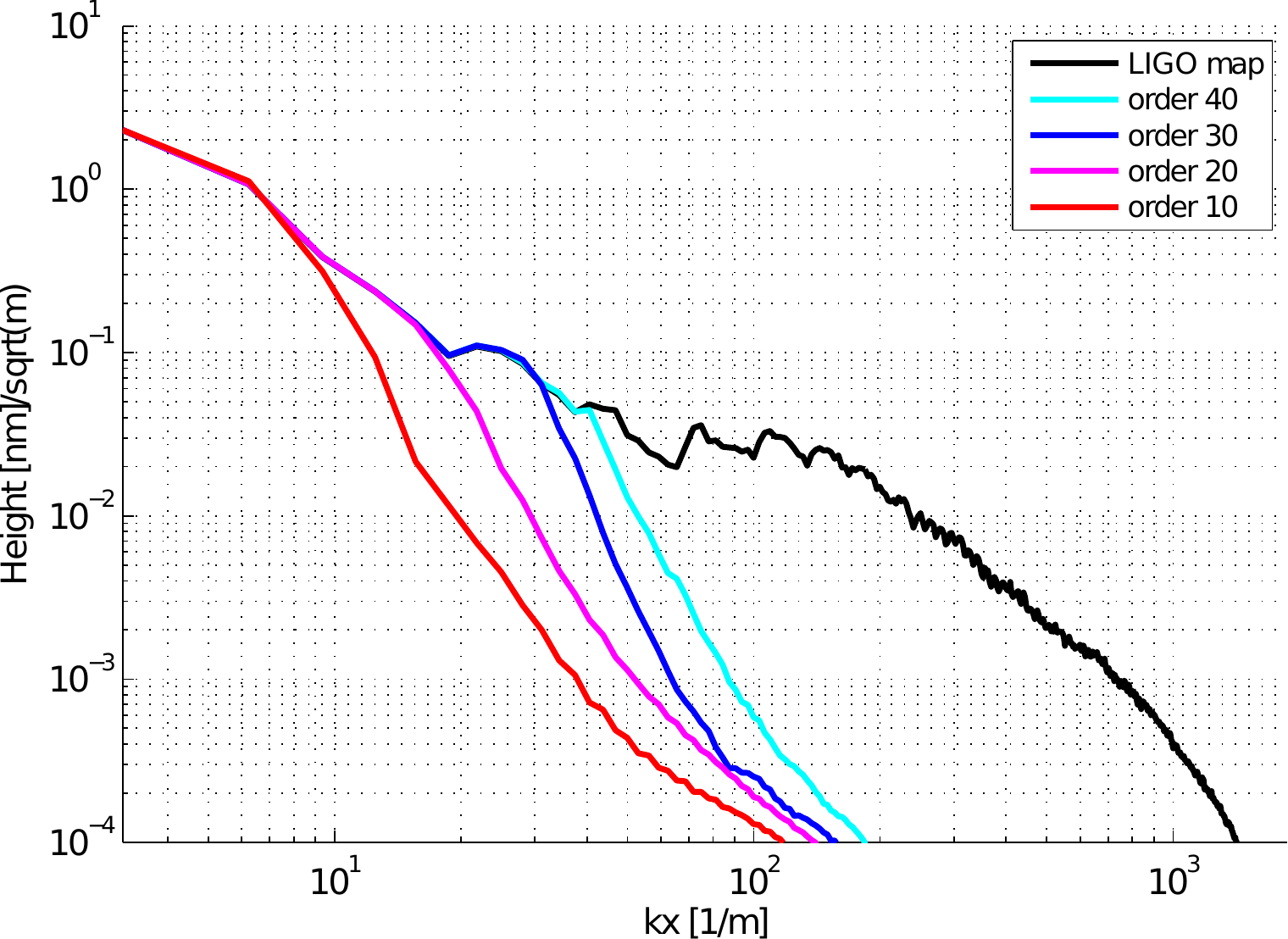} 
   \caption{A plot showing the spectrum of the Advanced LIGO mirror map, ETM08 and the spectra of maps created from Zernike polynomials present in ETM08.  The Zernike maps go up to a certain maximum order, recreating higher spatial frequencies with higher orders.}
   \label{fig:map_spectra}
\end{figure}

In the Zernike approach we look at the different polynomials present in mirror surface distortions.  This can be thought of as equivalent to looking at the spectra of the mirror surfaces as the different polynomials represent different spatial frequencies.  The plot in Fig.~\ref{fig:map_spectra} illustrates this.  The spectrum of the ETM08 Advanced LIGO mirror map is shown, along with the spectra of maps made up from the Zernike polynomials present in the ETM08 mirror.  Each of the Zernike maps recreates the LIGO map with polynomials up to a certain order.  The plots show that as the order of Zernike polynomials present increases the higher spatial frequencies are represented in the mirror map.  This is because these higher order polynomials represent the higher order spatial frequencies.  Looking at the spatial frequencies present in the Zernike polynomials we found that the frequencies depended on the order, $n$.  A consequence of this is that if we just consider the spatial frequencies present in the mirror maps we will not be able to distinguish between polynomials with different $m$.  As we have seen, the azimuthal index is very significant as it determines which modes are largely coupled from LG$_{33}$.  Therefore looking at the spatial frequencies doesn't identify the important shapes in the mirror surfaces.  The Zernike approach would seem to be the most suitable as this allows us to identify the interesting polynomials and modes. 

\end{document}